\documentclass[twocolumn,times]{aastex61}
\usepackage{multirow}
\usepackage{natbib}
\shorttitle{TiO WASP-33b draft}
\shortauthors{Nugroho et al.}
\begin{document}

\title{High-Resolution Spectroscopic Detection of TiO and Stratosphere in the Day-side of WASP-33b}

\correspondingauthor{Stevanus K. Nugroho}
\email{sknugroho@astr.tohoku.ac.jp}

\author{Stevanus K. Nugroho}\altaffiliation{Indonesia Endowment Fund for Education Scholar}\affil{Astronomical Institute, Tohoku University, Sendai 980-8578, Japan}
\affiliation{Department of Earth and Planetary Science, The University of Tokyo, Tokyo 113-0033, Japan}

\author{Hajime Kawahara}
\affiliation{Department of Earth and Planetary Science, The University of Tokyo, Tokyo 113-0033, Japan}
\affiliation{Research Center for the Early Universe, School of Science, The University of Tokyo, Tokyo 113-0033, Japan}

\author{Kento Masuda}
\affiliation{Department of Astrophysical Sciences, Princeton University, Princeton, NJ 08544, U.S.A}
\affiliation{NASA Sagan Fellow}

\author{Teruyuki Hirano}
\affiliation{Department of Earth and Planetary Sciences, Tokyo Institute of Technology, Tokyo 152-8551, Japan}

\author{Takayuki Kotani}
\affiliation{National Astronomical Observatory of Japan, Tokyo 181-8588, Japan}
\affiliation{Astrobiology Center, National Institutes of Natural Sciences, Tokyo, Japan}

\author{Akito Tajitsu}
\affiliation{Subaru Telescope, 650 N. A’ohoku Place, Hilo, HI 96720, U.S.A}

\begin{abstract}
We report high-resolution spectroscopic detection of TiO molecular signature in the day-side spectra of WASP-33 b, the second hottest known hot Jupiter. We used High-Dispersion Spectrograph (HDS; R $\sim$ 165,000) in the wavelength range of 0.62 -- 0.88 $\mu$m with the Subaru telescope to obtain the day-side spectra of WASP-33 b. We suppress and correct the systematic effects of the instrument, the telluric and stellar lines by using SYSREM algorithm after the selection of good orders based on Barnard star and other M-type stars. We detect a 4.8-$\sigma$ signal at an orbital velocity of $K_{p}$= +237.5 $^{+13.0}_{-5.0}$ km s$^{-1}$ and systemic velocity $V_{sys}$= -1.5 $^{+4.0} _{-10.5}$ km s$^{-1}$, which agree with the derived values from the previous analysis of primary transit.  Our detection with the temperature inversion model implies the existence of stratosphere in its atmosphere, however, we were unable to constrain the volume-mixing ratio of the detected TiO. We also measure the stellar radial velocity and use it to obtain a more stringent constraint on the orbital velocity, $K_{p} = 239.0^{+2.0}_{-1.0}$ km s$^{-1}$. Our results demonstrate that high-dispersion spectroscopy is a powerful tool to characterize the atmosphere of an exoplanet, even in the optical wavelength range, and show a promising potential in using and developing similar techniques with high-dispersion spectrograph on current 10m-class and future extremely large telescopes.

\end{abstract}

\section{Introduction} \label{sec:intro}
Thermal inversion in exoplanetary atmosphere is still a problem not completely understood in the theory of exoplanetary atmosphere. This inversion layer was predicted by \cite{2003ApJ...594.1011H} and \cite{2008ApJ...678.1419F} in a highly irradiated planet, which is caused by high-temperature absorber molecules such as TiO or VO that absorb UV and visible radiation from the incoming stellar radiation, and heat up the upper atmosphere. By measuring shallower eclipse depths at 4.5 $\mu$m and 5.6 $\mu$m, caused mainly by H$_{2}$O and CO, than the thermal continuum at the nearby band (3.6 and 9 $\mu$m) using the \textit{Spitzer Space Telescope} (\textit{Spitzer}), the first evidence of inversion layer was reported by \citet{Knutson2008ApJ...673..526K} in the atmosphere of HD 209458 b. Although there have been similar observations of secondary eclipse depth by the same telescope suggesting the detection of an inversion layer, it was shown by \citet{Hansen2014MNRAS.444.3632H} that several previous single-transit observations using \textit{Spitzer} have significantly higher uncertainties than previously known, which made the measured emission-like features doubtful. Using the CRyogenic high-resolution InfraRed Echelle SPectrograph \citep[CRIRES][]{Kaeufl2004SPIE.5492.1218K} in the Very Large Telescope (VLT), \citet{Schwarz2015A&A...576A.111S} observed the day-side of HD 209458 b and reported the non-detection of CO at 2.3 $\mu$m and constrained the nonexistence of the inversion layer in the pressure range of 10$^{-1}$--10$^{-3}$ bar, which is consistent with the results by \citet{2014ApJ...790...53Z} and \citet{Diamond2014ApJ...796...66D}. Despite the non-detection of CO in the day-side of HD 209458 b, \citet{Snellen2010Natur.465.1049S} detected a CO absorption feature at 5.6-$\sigma$ using a similar instrument in the transmission spectrum. One of the possible causes of the non-detection is that the average day-side atmosphere of HD 209458 b is near-isothermal in the pressure range that they probe, which makes the day-side spectrum almost featureless \citep{Schwarz2015A&A...576A.111S}.

\citet{Evans2017Natur.548...58E} reported the detection of H$_{2}$O emission features in a super-hot Jupiter, WASP-121b (T$_{eq}\sim$ 2500 K), using the combination of \textit{Hubble Space Telescope} (HST) and \textit{Spitzer}. The presence of water was resolved and detected at 5-$\sigma$ confidence level, strengthening the previous conclusion by \citet{Evans2016ApJ...822L...4E}, which made WASP-121b the first exoplanet with resolved emission features and detected stratosphere layer. Their findings also include the possible detection of VO in its atmosphere at the level of 1000x solar abundance, and a thermal inversion-like atmospheric structure, even though they only used a 1D atmospheric model structure and did not take the non-equilibrium chemistry into account. \citet{2017Natur.549..238S} observed WASP-19b during transit using a low-dispersion spectrograph, FORS2, on VLT. They confirmed the presence of water (7.9-$\sigma$) and simultaneously revealed the presence of TiO (7.7-$\sigma$), strongly scattering haze (7.4-$\sigma$) and sodium (3.4-$\sigma$). They also constrained the relative abundance of those molecules, however, the presence of a thermal inversion remains unproved because a transmission spectrum has less information on temperature structure of the atmosphere. Other evidence of thermal inversion has also been reported in the atmosphere of WASP-33 b (T$_{eq}\sim$ 2700 K), which is the second hottest hot Jupiter, by \citet{Haynes2015ApJ...806..146H}, who observed the day-side of the exoplanet using HST, and revealed a thermal excess at about 1.2 $\mu$m, which is consistent with the TiO spectral feature. Other than these, there has been no significant evidence, nor direct detection of TiO and/or VO in the atmosphere of hot Jupiters.

The non-detection of TiO or VO can be caused by several factors. \citet{2003ApJ...594.1011H} and \citet{Spiegel2009ApJ...699.1487S} suggested that gravitational settling could drag TiO/VO from the upper atmosphere to the colder layer in the deeper atmosphere. Meanwhile, owing to high-speed winds of the tidally locked hot Jupiter, condensed molecules can also be brought into the colder night side of the planet. If the temperature of the atmosphere is below the TiO/VO condensation level, it can condensate and if the vertical mixing rate, which is related to the temperature of the planet atmosphere is not high enough, it cannot be redistributed to the upper atmosphere. This effect is called the cold-trap effect. \citet{Knut2010ApJ...720.1569K} found a possible connection between the UV chromospheric stellar activity and the existence of an inversion layer in the atmosphere of known hot Jupiters. Hot Jupiters orbiting active stars tend to have no inversion layer and those orbiting quiet stars showed evidences of the existence of an inversion layer in their atmosphere. \citet{Knut2010ApJ...720.1569K} suggested that the increased UV intensity can probably destroy the compounds that are responsible of creating an inversion layer. Meanwhile, using high-dispersion spectroscopy, \citet{Hoej2015A&A...575A..20H} reported the inaccuracy of the TiO line list that was used at wavelengths shorter than 6300 $\AA$. The accuracy level, however, tends to increase at longer wavelengths, which was shown by the cross-correlation result between the spectrum of Barnard's Star and the model spectrum created by using the corresponding TiO line list \citep[see Figure 9 of][]{Hoej2015A&A...575A..20H}.

Recently, direct detection of molecular signature in exoplanet atmosphere using high-dispersion spectroscopy is one of the most widespread approach to the attempt of exoplanet characterization \citep[e.g][]{Snellen2010Natur.465.1049S, Crossfield2011ApJ...736..132C, Birkby2013MNRAS.436L..35B, deKok2013A&A...554A..82D, Hoej2015A&A...575A..20H, Schwarz2015A&A...576A.111S, Birbky2017AJ....153..138B, Esteves2017AJ....153..268E}. Unlike low-dispersion spectroscopy, high-dispersion spectroscopy can resolve molecular bands into individual absorption lines. The variation of Doppler shifts during observation caused by the orbital movement enables to distinguish absorption lines in the exoplanet spectrum from telluric lines, and ensures the unambiguous detection of specific molecules. Owing to these resolved individual lines it is also possible to investigate several physical parameters of the exoplanet such as the axial tilt \citep{Kawahara2012ApJ...760L..13K}, the projected equatorial rotational velocity \citep{Snellen2014Natur.509...63S}, wind speed \citep{Brogi2016ApJ...817..106B}, and the thermal inversion layer \citep{Schwarz2015A&A...576A.111S} of the planet. Obtaining the exoplanet spectrum can be done by cross-correlating the data with the exoplanet atmosphere spectrum model, after removing telluric and stellar lines using a specific method.

We observed WASP-33 b \citep{Smith2011MNRAS.416.2096S}, which is the second hottest known hot Jupiter (wavelength dependent brightness temperature of 3620 K), orbiting quiet $\delta$ Scuti stars. It has a retrograde orbit with a period of $\sim$ 1.22 days. It is an ideal choice for transmission spectroscopy measurements: owing to its unusually large radius \citep{Collier2010MNRAS.407..507C} and its high temperature making it not only suitable for secondary eclipse spectroscopy measurements, but also as the main target to find TiO/VO in its atmosphere, as the cold-trap effect is unlikely at this temperature level.

In this paper, we report a direct detection of TiO molecules and a stratosphere layer of WASP-33b, based on ground-based observation of the day-side emission spectrum in the visible wavelength range (6170-8817$\AA$). The observation, data reduction, and systematic effect removal (including the correction of the blaze function variation, common wavelength grid, and telluric and stellar line removal) are described in \S \ref{sec:obsndr}. We also examine our methods to create the model spectrum and to cross-correlate the data with the model spectrum. Here, we also explain the method to confirm the radial velocity of WASP-33 and the accuracy of TiO line list used by us. In \S \ref{sec:TiODetect}, the possibility of signal detection of TiO and the order-by-order (order-based) optimization of the SYSREM algorithm are explored. Here, we also show the final result after the optimization and the statistical test. This is followed by the discussion and the conclusions our findings in \S \ref{sec:discussandcon}.

\section{Observation and Data Reduction} \label{sec:obsndr}
\subsection{Subaru Observation of WASP 33}

\begin{figure}[t!]
\centering
\includegraphics[width=0.4\textwidth]{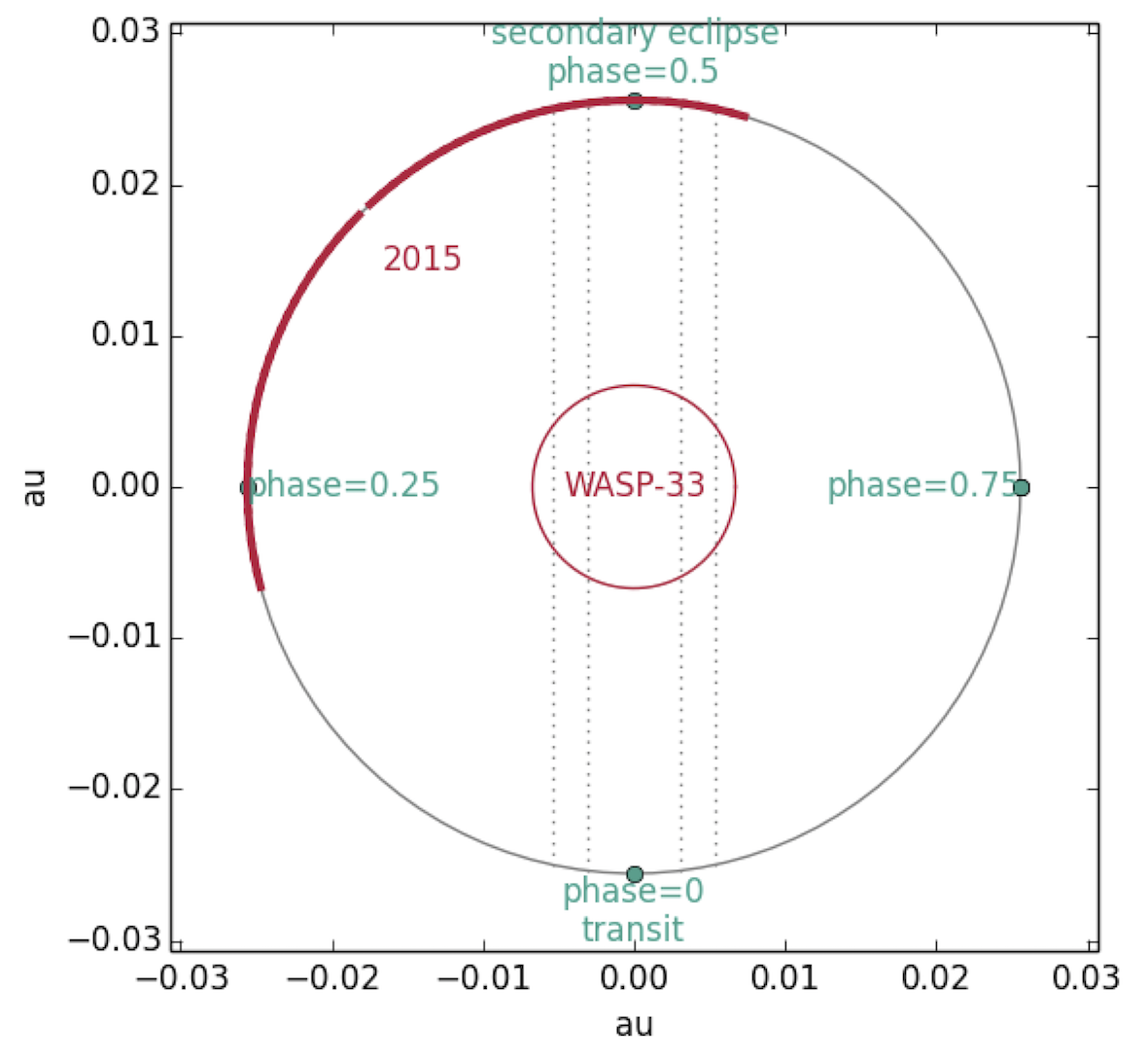}
\caption{The coverage of WASP-33 b orbital phase during our observation (showed by bold red line). The vertical dotted line shows the ingress and egress phase \label{fig:orbitalphase}}
\end{figure}

We observed WASP-33 on UT October 26 2015 (see Table \ref{tab:table1} for stellar and exoplanet physical and dynamical parameters) using High Dispersion Spectrograph \citep[HDS][]{Noguchi2002PASJ...54..855N} at f/12.71 optical Nasmyth focus of the the Subaru 8.2 m telescope (proposal ID: S15B-090, PI: H. Kawahara). The observation was conducted in a standard NIRc set up (without Iodine cell) with Messia5, 2x1 binning setting, and without image rotator. Image slicer 3 \citep{Tajitsu2012PASJ...64...77T}, each with slit width= 0.$\arcsec$2 was used, resulting in the highest spectral resolution of R= 165,000 (1.8 km s$^{-1}$ resolution), which were sampled by two detectors (blue and red CCDs) with 4100 $\times$ 2048 pixels (0.9 km s$^{-1}$ per pixel) containing 18 orders covering 6170 -- 7402 $\AA$, and 12 orders covering 7537 -- 8817 $\AA$, respectively. We obtained 52 spectra of WASP 33, each with an exposure time of 600 s, from air mass 2.97 to 1.96 on the other meridian covering 0.23 -- 0.56 exoplanet orbital phases (see Table \ref{tab:table1} for ephemeris and derived the orbital period, the orbital coverage of our observation is shown by bold red line in Figure \ref{fig:orbitalphase}) with a typical seeing of $\sim$ 0.$\arcsec$6 -- 0.$\arcsec$7.

\begin{deluxetable}{lcc}
\tablewidth{0pt}
\tablecolumns{2}
\tablecaption{System parameters of WASP-33  \label{tab:table1}}
\tablehead{\colhead{Parameter} & \colhead{Value}}
\startdata
\textbf{WASP-33} \\
$M_{\star}$ ($M_{\bigodot}$)    & 1.561$^{+0.045}_{-0.079}$\tablenotemark{a}  \\
                            & 1.512 $\pm$ 0.04 \tablenotemark{d} \\
                            & 1.495 $\pm$ 0.031 \tablenotemark{b} \\
$R_{\star}$ ($R_{\bigodot}$)    & 1.509$^{+0.016}_{-0.027}$  \\
$L_{\star}$ ($L_{\bigodot}$)    & 6.17 $\pm$ 0.43 \\
Spectral type & A5  \\
$T_{eff}$ (K) & 7400 $\pm$ 200\tablenotemark{b}\\
$\log$ $g$ & 4.3 $\pm$ 0.2 \\
$[$Fe/H$]$ & 0.1 $\pm$ 0.2\tablenotemark{b}  \\ 
$[$M/H$]$ & 0.1 $\pm$ 0.2\tablenotemark{a} \\  
Centre-of-mass velocity (km s$^{-1}$)&-2.19 $\pm$ 0.09\tablenotemark{b}  \\ 
                                 & -3.69 $\pm$ 0.09\tablenotemark{b}  \\ 
                                 & -2.11 $\pm$ 0.05\tablenotemark{b}  \\ 
$\textit{v}_{rot}$ sin $\textit{i}_{\star}$ (km s$^{-1}$)& 86.63$^{+0.37}_{-0.32}$\tablenotemark{c} \\
$d$ ($pc$)                         & 117 $\pm$ 2  \\
\hline
\textbf{WASP-33b} \\
$\textit{T}_{o}$-2450000 (BJD)     & 6934.77146 $\pm$ 0.00059\tablenotemark{c}  \\
$P$ (days)                         & 1.2198709\tablenotemark{a}  \\ 
$T_{transit}$ (days)             & 0.1143 $\pm$ 0.0002\tablenotemark{a}  \\ 
$T_{ingress}$ (days)                & 0.0124 $\pm$ 0.0002\tablenotemark{a}  \\ 
$a/R_{\star}$                     & 3.69 $\pm$ 0.01\tablenotemark{a}  \\ 
$\textit{i}$ ($^{\circ}$)            & 88.695$^{+0.031}_{-0.029}$\tablenotemark{c}\\ 
 2015
$M_{P}$ ($M_{J}$)                & 3.266 $\pm$ 0.726\tablenotemark{a}  \\
$R_{P}$ ($R_{J}$)                 & 1.679$^{+0.019}_{-0.030}$\tablenotemark{a}  \\
$\log$ $g_{P}$ $[$CGS$]$             & 3.46$^{+0.08}_{-0.12}$\tablenotemark{a} \\
$K_{p}$ (km s$^{-1}$)            & 231.11 $^{+2.20}_{-3.97}$\tablenotemark{a}  \\
                                & 228.67 $^{+2.00}_{-2.04}$\tablenotemark{d}  \\
                                & 227.81 $^{+1.56}_{-31.59}$\tablenotemark{b}  \\
\enddata
\tablenotetext{a}{Adopted from Kov$\acute{a}$cs et al. 2013} 
\tablenotetext{b}{Adopted from \cite{Collier2010MNRAS.407..507C}}
\tablenotetext{c}{Adopted from \cite{Johnson2015ApJ...810L..23J}}
\tablenotetext{d}{Adopted from \cite{Smith2011MNRAS.416.2096S}}

\end{deluxetable}

\subsection{M-dwarfs Spectra}
\citet{Hoej2015A&A...575A..20H} showed that the TiO line list used by them is not accurate at short wavelengths ($<$ 6000 $\AA$) however, it tends to be more accurate at longer wavelengths. For a robustness of analysis, we checked the accuracy of the TiO line list we use in this paper, by comparing it with the spectra of Barnard's Star (M4V) and HD 95735 (M1.5V), as TiO is commonly found in the atmosphere of M-type stars and brown dwarfs \citep{Burrows1999ApJ...512..843B, Burrows2001RvMP...73..719B}. In 2016, we observed Gl752A (M3V), and HD173739 (M3V) with the Subaru telescope using the same instrumental configuration as in 2015 (proposal ID: S16A-107, PI: H. Kawahara). To check the robustness of our cross-correlation we also downloaded and used the calibrated spectrum of Proxima Centauri from the ESO Science Archive Facility. The spectrum was obtained by XSHOOTER at VLT between 5336.6 $\AA$ and 10200 $\AA$ with a resolution of R= 18,000 (proposal ID: 092.D-0300(A), PI: Neves). 
\begin{figure*}
\gridline{\fig{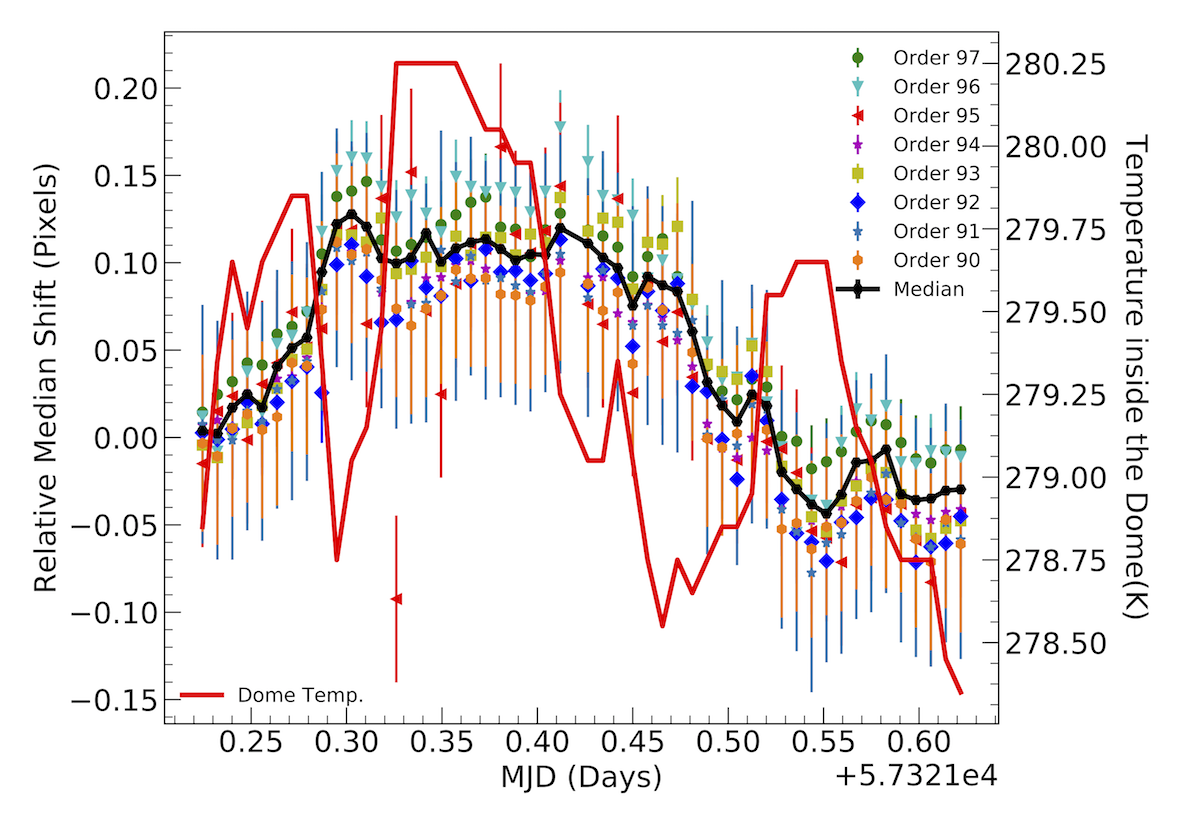}{0.48\textwidth}{(a)} 
         \fig{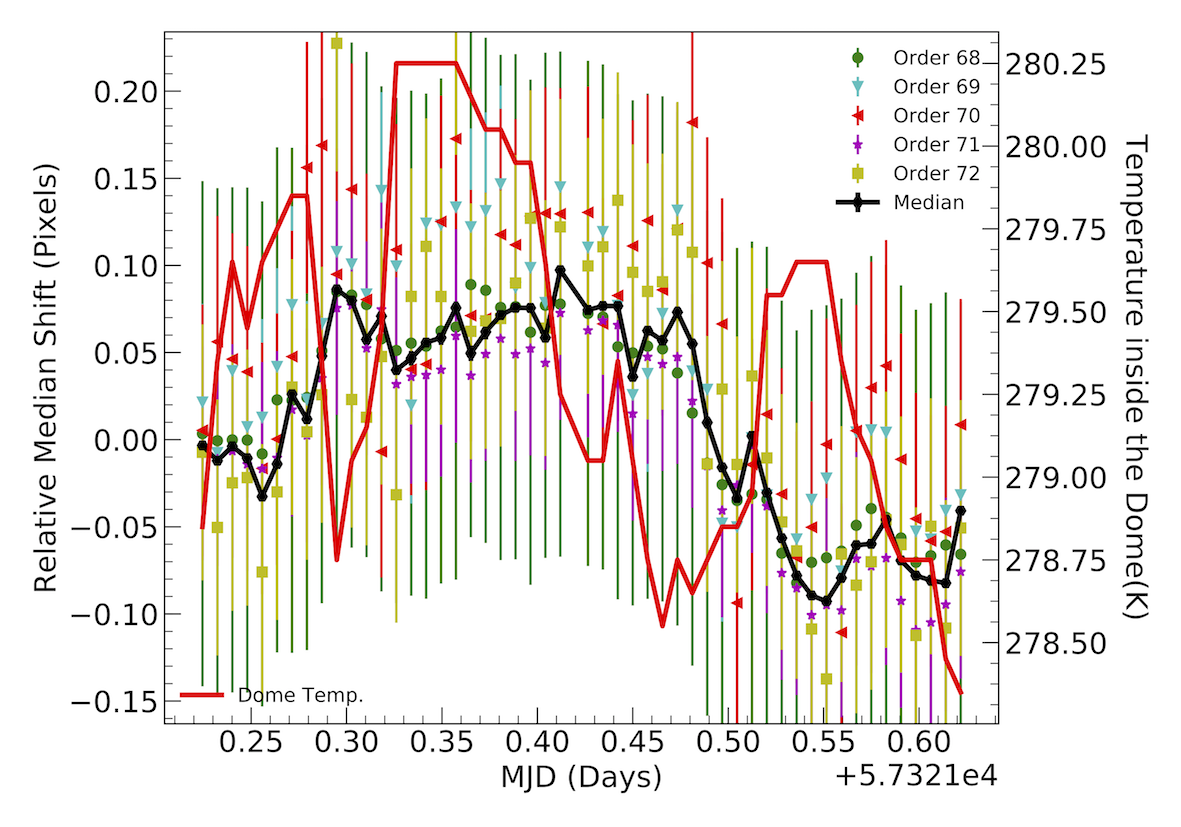}{0.48\textwidth}{(b)}}
\caption{Relative shifts of telluric lines compared to those in the first frame in the blue CCD (1) and red CCD (2). Noticeably the general trend of the shift follows the trend of temperature variation inside the dome (solid red line)\label{fig:figshift}}
\end{figure*}

\subsection{Standard Reduction}
The data were reduced using IRAF tools\footnote{The Image Reduction and Analysis Facility (IRAF) is distributed by the US National Optical Astronomy Observatories, operated by the Association of Universities for Research in Astronomy, Inc., under a cooperative agreement with the National Science Foundation.} and a custom-built application written in Python 2.7. We corrected the over-scan and non-linearity using CL scripts that can be obtained from the HDS website\footnote{\url{http://www.subarutelescope.org/Observing/Instruments/HDS/index.html}} before debiasing the frames. During the analysis of \citet{Narita2005PASJ...57..471N} data, \citet{2008A&A...487..357S} noticed the non-linearity effect in HDS. After fixing this problem empirically, the analysis gave the detection of sodium in the atmosphere of HD 209458b. The CCD response (gain) for high signal-to-noise (S/N) spectra is higher than those in a lower S/N spectra, which would make the correction of telluric and stellar lines difficult. This problem was solved by \citet{2010PNAOJ..13....1T}, who provided a CL script to correct this effect after applying the over-scan correction. The scattered light in the inter-order area was fitted with a cubic spline function along the slit and dispersion direction individually, using $\textit{apscatter}$ for the arc lamp frame, smoothed and subtracted from all science frames. A median flat frame was calculated from 71 dome-flat frames to correct the pixel-to-pixel sensitivity variation, and normalized using the $\textit{apnormalize}$ task in IRAF in order to conserve the fringe pattern along the slit direction. Using the CL script $\textit{hdsis\_ecf.cl}$ \footnote{\url{https://www.naoj.org/Observing/Instruments/HDS/hdsql/hdsql-cl-20170807.tar.gz}}, we flat-fielded and extracted 1-dimensional spectra of total 30 orders. Using $\textit{hdsis\_ecf.cl}$ the science spectra were sliced along the slit from $-$12 to $+$8 pixels and from $-$11 to $+$8 pixels relative to the center of the order tracer of each order for the blue and red CCD respectively, divided them by the slice of the normalized median flat frame and sum combined. If a larger width was used, a high-frequency noise would appear along the wavelength in the extracted spectra, caused by the edge of the aperture, which was defined earlier in $\textit{apnormalize}$ by using a median flat frame. We derived the wavelength solution by identifying 133 emission lines of Thorium-Argon arc lamp frames taken at the beginning and the end of the observation, and fitted with a fifth and third order Chebyshev function corresponding to the dispersion and slit directions respectively, using \textit{ecidentify}. The pixel RMS value of the residual fitting was $\sim$ 0.0012 for both CCDs. Then, the spectra were assigned by interpolating between the preceding and the following Thorium-Argon arc lamp frames in respect of the observation time, using \textit{refspec} with the relative distance of the spectra to the reference spectra as the weight of the interpolation. Then, the wavelength solution was applied using \textit{dispcor}. These steps were applied to the non-normalized median flat frame to create the corrector for the blaze function. All science frames were then divided by the blaze function corrector to create the final reduced frames. Then the next step is correcting the variation of blaze function along the time which is given in detailed in Appendix \ref{Appendix_A}.

\subsection{Common Wavelength Grid}\label{subsec:commonwave}
We measured the shifts of the spectrum during the observation, by measuring the relative shifts of strong telluric lines to the reference spectrum. First, we detected telluric lines using \textit{peakutils}\footnote{\url{https://bitbucket.org/lucashnegri}} (15 -- 20 lines on average per selected order) in the order that contain well-spread telluric lines in the wavelength direction. We then fitted a Gaussian function to determine the precise centroid of each line and compared it with the centroid of the same lines in the first frame. The relative shift of each order was calculated by taking the median combination of the shifts in that order. There are order dependent shifts of about 0.05 pixel, but because the information for the order that does not has strong telluric lines is not available and the value is too small to affect the result (0.05 pixels corresponding to $\sim$0.04 km s$^{-1}$), we decided to ignore it. The results of all selected order were combined to estimate the shift of the corresponding frame, by calculating its weighted median combined. 

The amplitude of relative shifts is about 0.13 and 0.08 pixel ($\sim$0.12 and 0.07 km s$^{-1}$) for blue and red CCD, respectively. Noticeably the general trend of the shifts are correlated with the shifts of the temperature inside the dome ($\Delta$T $\sim$1.75 K) during the observation (see Figure \ref{fig:figshift} a $\&$ b). This wavelength-temperature shift relation trend is relatively weaker than those for the CRIRES in VLT reported by \citet{Brogi2013ApJ...767...27B} (1.5 K change in temperatures corresponding to 1.5 pixels shifts), which gave a non-detection of the 51 Peg b signal due to the odd-even effect for detector 4. 

Then, the spectra were shifted with spline interpolation, based on the weighted median combined shift of each frame, and then re-sampled into a common wavelength bin. The spectra of each order were then stacked in a matrix with wavelength bin values in its column, and spectrum numbers (or time/orbital phases) in the rows. 5-$\sigma$ clipping was performed for each wavelength bin to identify any bad pixels/cosmic rays, which are replaced by the mean value of the bin. We excluded these bad region for the rest of the analysis, which is 5.30\% and 7.19\% of all pixels in blue and red CCDs, respectively, mostly because they are on the edge of the CCD.

\begin{figure*}
\gridline{\fig{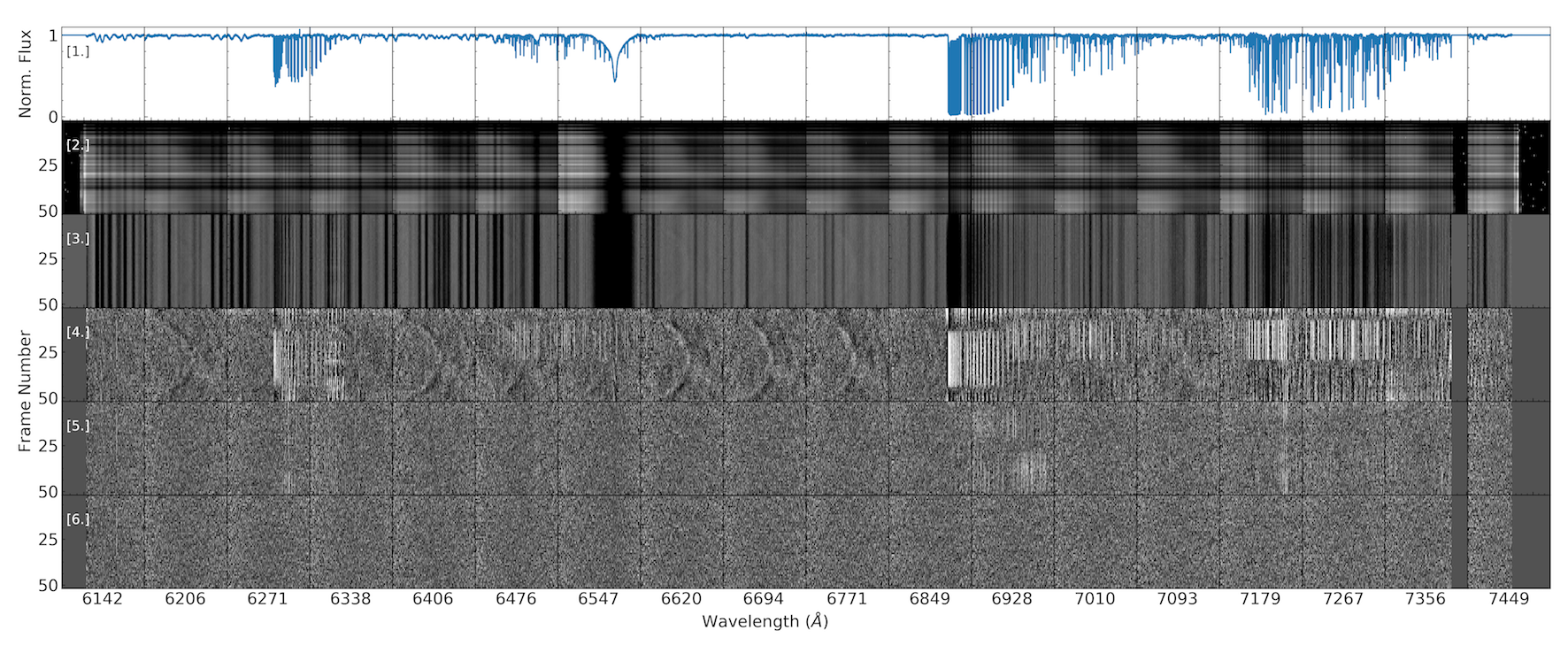}{1.\textwidth}{(a)}}
\gridline{\fig{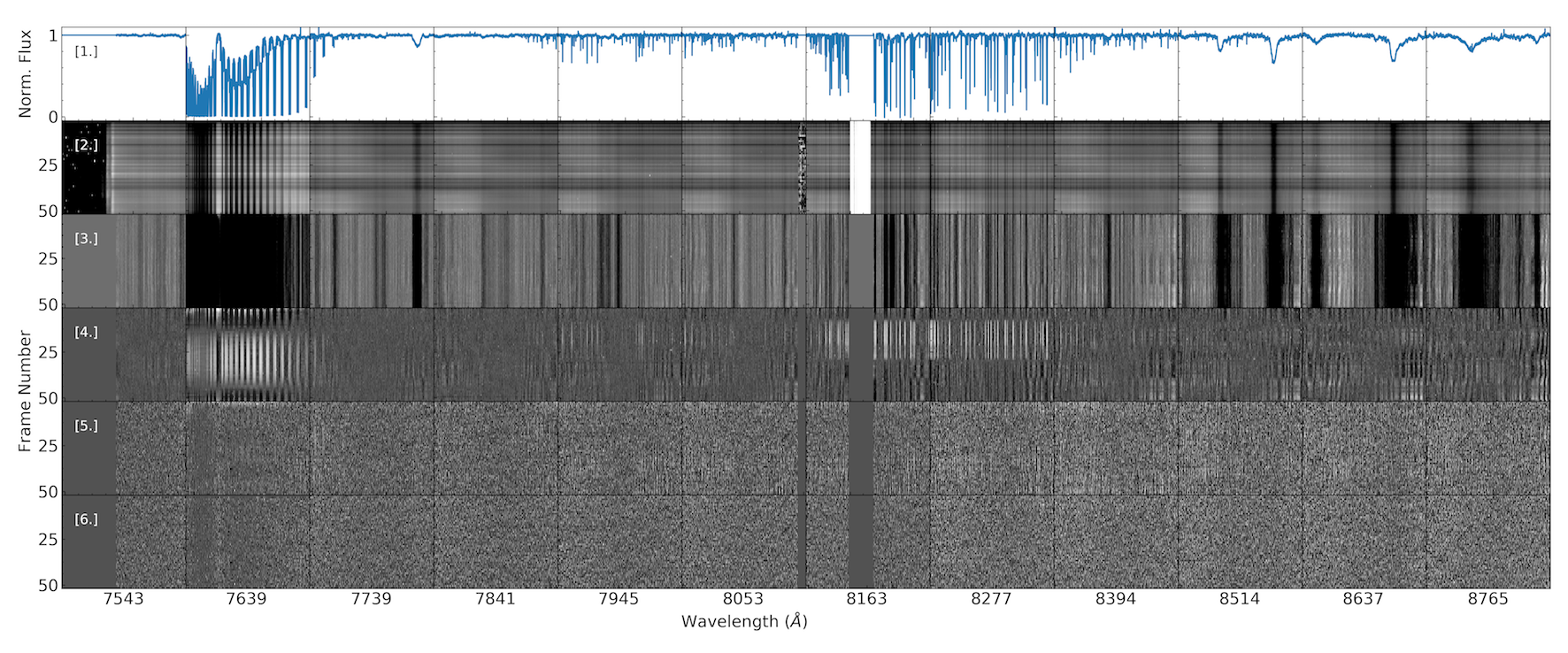}{1.\textwidth}{(b)}}
\caption{Reduction process for each order of both blue (a) and red CCDs (b). The first row ([1]) shows an example of the normalized 1D spectrum of WASP-33 following the correction of the blaze function variation and the common wavelength grid iteration. The next row ([2]) shows the 2D spectrum with the wavelength as the horizontal axis, each label representing the median wavelength of the order, while the vertical axis is the frame number. The third row ([3]) shows the final reduced spectra after the correction of the blaze function in common wavelength grid. The variation of brightness along the frames for all orders is due to the blaze function variation. The fourth row ([4]) shows the mean subtracted spectra as the input to SYSREM. The fifth and sixth rows ([5] $\&$ [6]) show the residual spectra after running SYSREM with 1 and 4 iteration and through the double high-pass filter, at the latter stage almost all telluric lines have been removed. The masked bad regions are shown by grey areas in all rows except the first one. \label{fig:figred}}
\end{figure*}

\subsection{Removal of Telluric and Stellar Absorption Lines \label{ss:tmovt}}
In the studied wavelength range, telluric and stellar absorption lines are dominating the spectrum, while the exoplanet to stellar flux contrast is expected at the level of 10$^{-3}$ \citep[when extrapolated from the best-fitted spectrum in][]{Haynes2015ApJ...806..146H}. Removal of telluric and stellar lines is critical in order to detect the TiO signature by cross-correlation, as the exoplanet spectrum is buried under sharp noises due to those lines. The the spectrum of WASP-33b is expected to be Doppler-shifted from +230 km s$^{-1}$ (at its maximum) to $-$54 km s$^{-1}$ (at the end of the observation), while telluric and stellar lines remain relatively stationary during the course of the observation. Telluric lines, which are dominated mostly by water and oxygen lines, are varied in strength due to the changes of geometric air mass and water vapor column of the atmosphere. 

The quasi-static telluric and stellar absorption lines were removed by implementing the SYSREM algorithm to remove systematic effects (variation of atmospheric condition, the changing of CCD efficiency, variation of point spread function, etc.) without any priors in a large set of light curves \citep{Tamuz2005MNRAS.356.1466T, Mazeh2007ASPC..366..119M}. It has been used either in the detrending/systematic effects removal of transit surveys \citep[e.g SuperWASP, CoRoT light curves][]{Collier2006MNRAS.373..799C, Ofir2010MNRAS.404L..99O}, or in the removal of quasi-static telluric lines in the similar analysis to this work \citep[e.g.][]{Birkby2013MNRAS.436L..35B, Birbky2017AJ....153..138B, Esteves2017AJ....153..268E}. Each wavelength bin (4100 wavelength bin per order on average) is treated as a ``light-curve'' consisting of 52 frames (including the frames when WASP-33 b in the secondary eclipse phase). Then each ``light-curve'' was subtracted from their mean value before applying the SYSREM algorithm order-by-order (fourth row in Figure \ref{fig:figred}). The detailed description of the SYSREM algorithm is given in Appendix \ref{Appendix_B}.

The step-by-step removal of telluric can be seen in Figure \ref{fig:figred}. It can be noticed that there are curve shaped features, which can be seen in the blue CCD spectra matrix, and the degree of the curvature decreases as increasing wavelength. These features can be caused by the imperfection of the correction of the blaze function variation. As the residual was cross-correlated with the high-frequency TiO features, this feature does not affect the results and these were roughly removed after a double high-pass filter was applied (see Figure \ref{fig:figred}).

Any linear systematic variation along all wavelength bins in each order can be found and removed from the spectrum starting with the most significant one, such as air mass variation. Ideally, the expected residual is the Doppler-shifted exoplanet spectrum only, with additional noise. We perform two cases of the SYSREM reduction: One is that we remove the first ten systematics (henceforth SYSREM iteration, $N_{\mathrm{sys}} = 10$) for all of the orders (SYSREM with common iteration number in \S \ref{ss:common}). This simple procedure gives a conservative estimate of the signal detection. Another attempt is that we determine the optimum number of subsequent SYSREM iterations for each order that gives the optimum systematic removal (Order-based SYSREM optimization in \S \ref{subsec:sysopti}). The latter is based on the fact that SYSREM tends to remove the exoplanet spectrum after the Doppler-shifted variation dominate the systematic change in the spectrum. 

Before performing cross-correlation for each residual, we applied a double high-pass filter similarly to the case of M-dwarfs star spectrum, using a smoothing function with a 25-pixel width for the first filter and 51-pixel width for the second filter, to remove any low-frequency variations along the spectrum, before cross-correlating with a grid of the Doppler-shifted WASP-33b model spectrum. Then, the values of each wavelength bin were weighted by their noise, which is defined by the standard deviation of the bin as a function of time. The results of this removal process are 10 sets of smoothed weighted SYSREM residuals for each CCD.

\subsection{Spectral Template}\label{sec:atmos}
\begin{figure*}[hb]
 \centering
\includegraphics[width=0.9\textwidth]{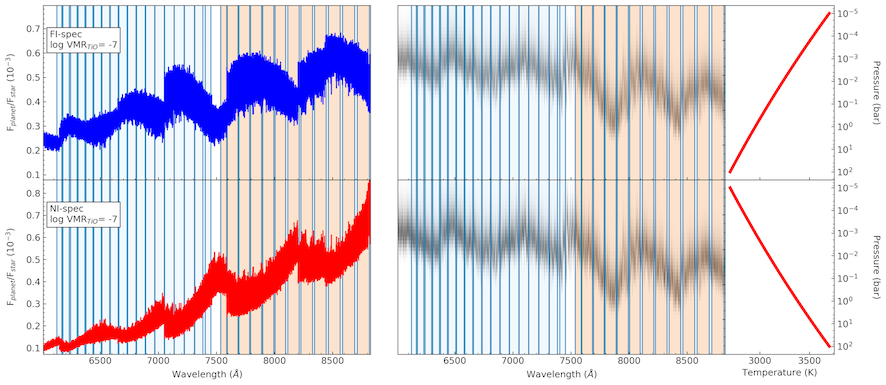}
\caption{WASP-33b spectrum model using full inversion (FI) and no inversion (NI) of the TP profile. The left panel shows the WASP-33b model spectrum for both FI-spec and NI-spec. The color shade represents the wavelength range of the observed order, blue, and orange color represent the wavelength range in the blue and red CCDs, respectively. The vertical axis shows the planet to star flux contrast in the level of 10$^{-3}$. The middle panel shows the weight function along the wavelength for the pressure range considered. The right panel shows our adopted temperature profile of the WASP-33b atmosphere. \label{fig:figWASP33bmodelalno}}
\end{figure*}

\begin{figure*}[hb]
 \centering
\includegraphics[width=0.9\textwidth]{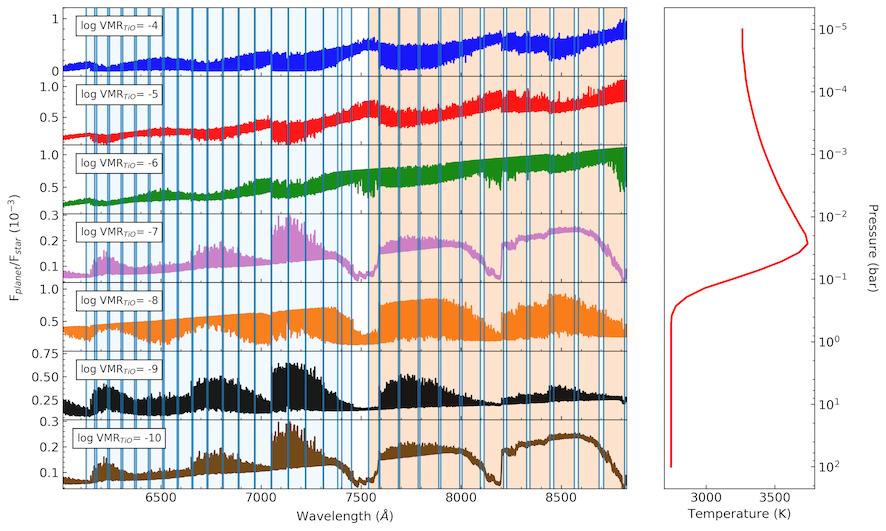}
\caption{WASP-33b spectrum model using TP profile of \citet{Haynes2015ApJ...806..146H}. The left panel shows the WASP-33b model spectrum of various VMR. The vertical axis shows the planet to star flux contrast in the level of 10$^{-3}$. The right panel shows our adopted temperature profile of the WASP-33b atmosphere.\label{fig:figWASP33bmodelhay}} \end{figure*}

To constrain the existence of TiO molecules in WASP-33 b emission spectra by cross-correlation, model spectra was generated with various profiles and volume mixing ratio (VMR) of TiO, assuming several atmospheric models as described in Table \ref{tab:table_specmodel}. Three different temperature--pressure (T/P) profiles were adopted, which described the average vertical temperature structure of the planetary day-side atmosphere: full inversion (FI), non-inversion (NI) as shown in the right panels of Figure \ref{fig:figWASP33bmodelalno}, and T/P profile from \citet{Haynes2015ApJ...806..146H}, known as the H-model, as shown in the right panel of Figure \ref{fig:figWASP33bmodelhay}. For the NI model, it was assumed that the temperature decreases with a constant lapse rate from P$_{0}$= 10$^{2}$ bar with T$_{0}$= 3700 K to P$_{1}$= 10$^{-5}$ bar with T$_{1}$= 2700 K. For the FI model, the temperature increases from P$_{0}$= 10$^{2}$ with T$_{0}$= 2700 K to P$_{1}$= 10$^{-5}$ with T$_{1}$= 3700 K. For both models, a constant TiO concentration at solar abundance level was assumed, that is, VMR= $10^{-7.2}$ and no other molecule in the atmosphere. For the H-model, seven constant VMRs of TiO from sub-solar to super-solar abundance level were assumed, $\log$ VMR$_{TiO}$= [$-$4, $-$5, $-$6, $-$7, $-$8, $-$9, $-$10]. The atmosphere was divided into 50 layers, which were evenly spaced in log pressure between 10$^{2}$ to 10$^{-5}$ bar, and the altitudes were calculated by assuming hydro-static H$_{2}$ dominant atmosphere and using the derived mass and a radius of WASP-33b from the reference (Table \ref{tab:table1}). We also modeled the non-inverted atmosphere (henceforth M-dwarf model) with T$_{0}$= 2600 K at P$_{0}$= 0.01 bar and T$_{1}$= 4000 K at P$_{1}$= 1 bar assuming a constant rate of temperature change with $\log$ pressure, resulting in TiO absorption features only for $\log$ VMR= $-$7.

\begin{table}[htb]
\tablewidth{0pt}
\caption{Models of planetary atmosphere}
\label{tab:table_specmodel}
\begin{tabular}{ccc}
\hline
\hline
Name & T/P Profile & log VMR  \\
\hline
\textbf{FI} & Full inversion & $-$7 \\
\textbf{NI} & No inversion & $-$7 \\
\textbf{H} & Realistic inversion (\cite{Haynes2015ApJ...806..146H}) & $-$4 -- $-$10 \\
\hline
\end{tabular}
\end{table}

\begin{figure*}[t!]
 \centering
\includegraphics[width=0.9\textwidth]{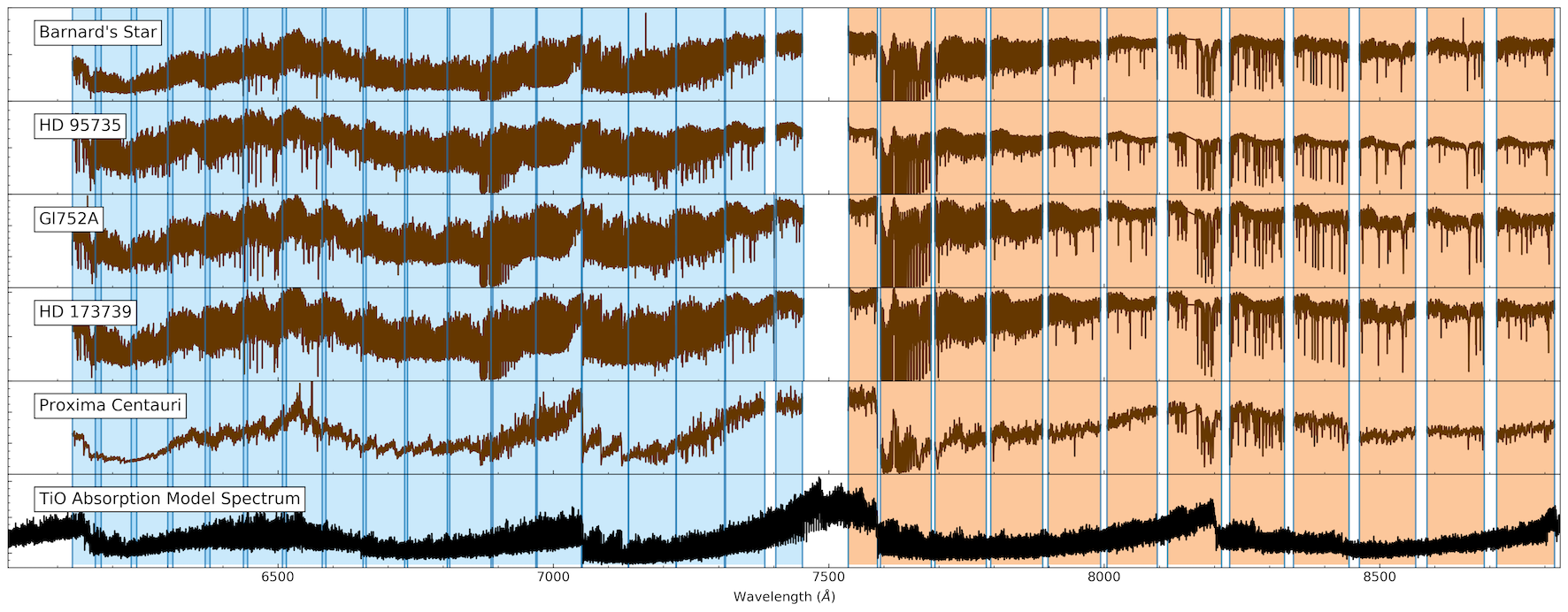}
\caption{Observed spectrum of five M-dwarf stars that were used for the TiO line list accuracy comparison. The bumps on the 4 first spectra are caused by the uncorrected blaze function. Before order-by-order cross-correlation, the spectrum was normalized by its continuum profile. The color shade represents the wavelength range of the observed order, the blue color is in the blue CCD and orange color is in the red CCD. \label{fig:figMdwarfs}} 
\end{figure*}

The cross section of molecules was calculated using \textit{Python scripts for Computational Atmospheric Spectroscopy} (Py4CATS\footnote{see \url{http://www.libradtran.org} for details}). The procedure of computing the cross section is described in Appendix \ref{Appendix_C}. Then, the cross-sections were combined into continuum absorption coefficients and integrated along the line-of-sight through the atmosphere, resulting delta optical depths per layer. 
The Schwarzschild equation was solved by integrating the Planck function versus the monochromatic transmission, along the line-of-sight and convolved with a Gaussian kernel to match with the HDS resolution.  In total, we produced 9 WASP-33b mock spectra and 1 non-inverted model for TiO line list accuracy analysis; 1 full inversion (FI-spec), 1 no inversion (NI-spec), 7 spectra for Haynes (H-spec), and 1 M-dwarf model. See Figure \ref{fig:figWASP33bmodelalno} for FI-spec and NI-spec, Figure \ref{fig:figWASP33bmodelhay} for H-spec, and Figure \ref{fig:figMdwarfs} for 5 M-dwarfs spectra$+$1 M-dwarf model spectrum.

In this analysis, the systemic velocity (V$_{\text{sys}}$) is one of the important parameters to confirm the detection (if there is any); thus by measuring radial-velocity (RV) of WASP-33 and comparing it with the previous results, the confidence and the robustness of our analysis are improved. Instead of creating the WASP-33 comparison spectra by our self, we used the stellar model spectrum from \citet{2014MNRAS.440.1027C} for T$_{\text{eff}}$= 7500 K, $\log$ g= 4.5, $[$Fe/H$]$= $+$0.2 and $[\alpha$/Fe$]$= 0. The stellar model spectrum was convolved to the HDS resolution and rotationally broadened to resemble the Doppler broadening caused by its fast rotation (henceforth Coelho model spectrum).

\begin{figure*}[t!]
\gridline{\fig{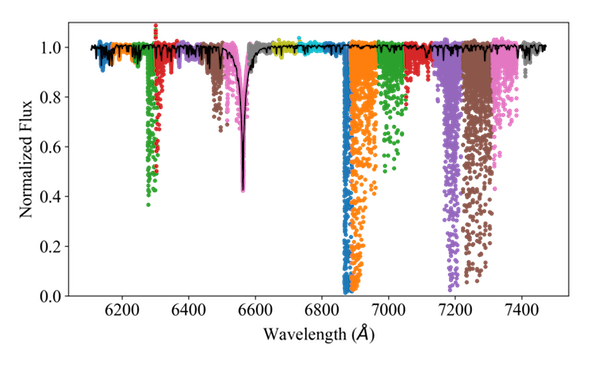}{0.45\textwidth}{(a)} 
         \fig{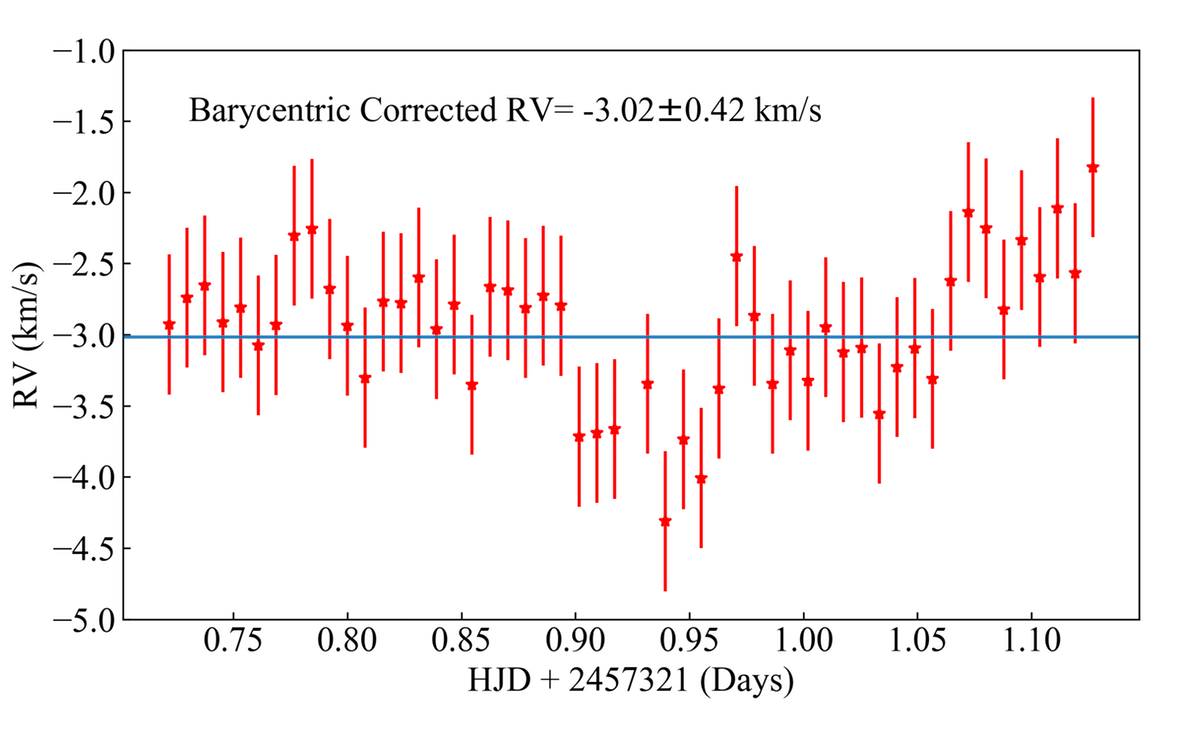}{0.45\textwidth}{(b)}}
\caption{(a) The colored line is an example of blue CCD spectrum of WASP-33 after standard reduction. Different colors mean different order. The black line is Coelho model spectrum. (b) The barycentric corrected RV of WASP-33 during the observation (red star), as a result of combining the RV of each order that has the highest cross-correlation signal. The blue line is the weighted median combine of all frames that was taken as the RV of WASP-33.The error bar is the 1-$\sigma$ scatter across the observation time. \label{fig:figRV}}
\end{figure*}

\subsection{System Velocity from Stellar Spectra}\label{subsection:wasp33rv}

To measure the radial velocity of the WASP-33 system, we analyzed the standard reduced WASP-33 spectra and masked the telluric lines of the selected order that contain significant stellar absorption lines. Then, the spectra were cross-correlated order-by-order with the Doppler-shifted Coelho model spectrum from $-$100 km s$^{-1}$ to $+$100 km s$^{-1}$ with 0.1 km s$^{-1}$ intervals. In this cross-correlation, only the spectra in blue CCD was used, because there are a lot of telluric lines and less stellar absorption lines in the red CCD spectrum. For each selected order, the RV value with the highest cross-correlation signal was extracted, and median combined to calculate the RV of WASP-33 for the corresponding frame. Then, the weighted median combine of the RV of all frames was calculated, following the correction of the barycentric radial velocity difference to have the final WASP-33 RV value.

The results can be seen in Figure \ref{fig:figRV}b, and the median of the measured RV is $-$3.02 $\pm$ 0.42 km s$^{-1}$. Although this value is different from the value $-$9.2 $\pm$ 2.8 km s$^{-1}$ \citep{Gont2006A&AT...25..145G} in the SIMBAD database, our result is consistent with \citet{Collier2010MNRAS.407..507C}, which used Doppler tomography technique to confirm the existence of the planet and measured the center-of-mass velocity (see Table \ref{tab:table1}). Therefore we used this value as the radial velocity of the WASP-33 system, which is taken into consideration when evaluating the result of the WASP-33b vs TiO model cross-correlation analysis.

\subsubsection{Line List Accuracy and Excluding Bad Orders} \label{tioaccuracy}
\begin{figure*}[t!]
\centering
\includegraphics[width=0.9\textwidth]{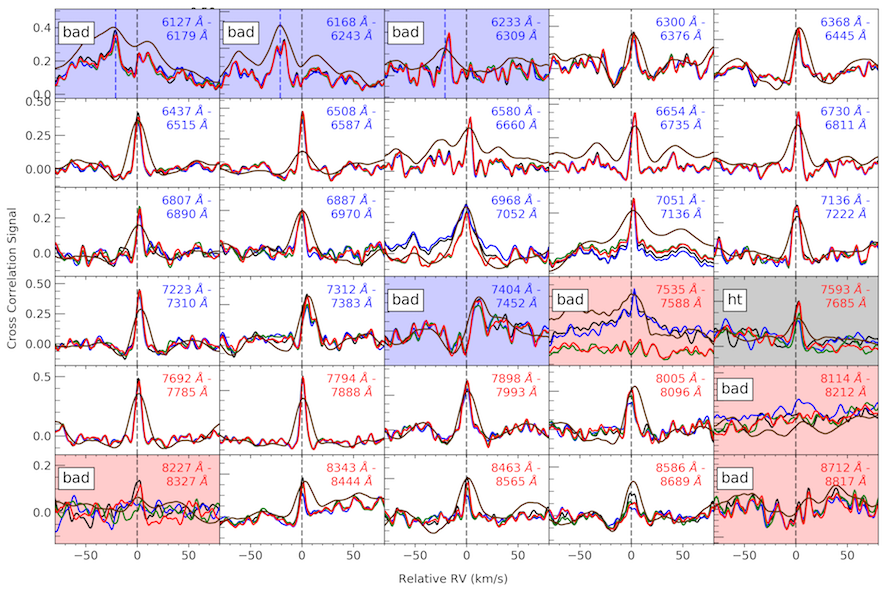}
\caption{Cross-correlation results between M-dwarf model and five M-dwarf spectra, Barnard's Star (black line), HD 95735 (blue line), Gl752A (green line), HD173739 (red line), and Proxima Centauri (brown line). The color of the label of the wavelength range also represents the order to which the CCD belongs. Note that the y-axis scale is not uniform, thus the orders cannot be compared. The black dash line is the rest-frame RV of each M-dwarf stars, while the vertical blue dashed line in the first three orders of the blue CCD is the position of CCF peaks on those orders. The blue and red shaded panel show the bad shape CCF that are masked for the rest of the analysis (bad). The grey shaded panel shows the masked order in the later analysis due to heavily contaminated by strong telluric lines (ht). \label{fig:figmdwarfsresult}}
\end{figure*}
To check the TiO line list accuracy, five standard reduced M-dwarf spectra were cross-correlated order-by-order with the M-dwarf model. The model spectra were Doppler-shifted from $-$80 km s$^{-1}$ to $+$80 km s$^{-1}$ relative to the RV of the expected target, from SIMBAD database with 1 km s$^{-1}$ intervals. The object and model spectrum was divided by their continuum profile, calculated by applying a double high-pass filter with 501 pixels and 1001 pixels of smoothing function before the cross-correlation, in order to maintain the cross-correlation scale (from $-$1 to 1). 

The cross-correlation results of five M-dwarf spectra versus the M-dwarf model is shown in Figure \ref{fig:figmdwarfsresult}. As \citet{Hoej2015A&A...575A..20H} expected, the accuracy of the TiO line list improved in the longer wavelength, although there are cross-correlation functions (CCFs) of several order, which blue-shifted from their expected radial velocity, and several others do not have any significant peak. For the other of the order, the peaks are located at their expected radial-velocity. The shifts are most likely caused by the inaccuracy of the TiO line list itself, because the first three orders of the blue CCD show similar blue-shifted CCF for the five different M-dwarf spectra. The shift is about $\sim$ 21 km s$^{-1}$ (shown by blue dashed line in Figure \ref{fig:figmdwarfsresult}). There are several orders that have no significant peak, which can be caused by the imperfection of our simple atmosphere modeling or the inaccuracy of the line list. Note that all orders that have no significant CCF peak, have a large bad region except the last order of the red CCD (see Figure \ref{fig:figred} and/or Figure \ref{fig:figmdwarfsresult}). These bad orders are excluded for the rest of the analysis including order 2 in red CCD due to heavily contaminated by strong telluric lines.

\section{TiO Signal Detection}\label{sec:TiODetect}
\subsection{Results from SYSREM with Common Iteration Number \label{ss:common}}
As described in \S \ref{ss:tmovt}, we first analyze the spectra with a common SYSREM iteration number for all orders. A grid of Doppler-shifted WASP-33b model spectrum was cross-correlated with weighted SYSREM residuals. The Doppler-shifted model spectrum covers planet radial velocity (RV$_{p}$) between $-$169.69 km s$^{-1}$ $\leq$ RV$_{p}$ $\leq$ $+$393.30 km s$^{-1}$ with 0.5 km s$^{-1}$ intervals, corresponding to half of the HDS sampling resolution. For each detector, the CCFs of every good orders (all orders excluding the bad orders explained in \S \ref{tioaccuracy}) were summed. This value is listed in the CCF matrix with a dimension of 1127 (RV as column) $ \times$ 52 (orbital phase as row). Then, the CCF matrix of blue and red CCDs was summed to calculate the final CCF matrix. These steps were done for all SYSREM residuals. 

We then calculate the CCF map in the $K_{p}$ -- $V_{sys}$ plane for all spectrum models by doing the following steps. The CCF of the frames (40 frames in total, excluding the frames when WASP-33b was expected in the secondary eclipse phase) was integrated along the expected RV$_{p}$ curve:
\begin{equation}\label{eq:10}
RV_{p} (t)= K_{p} \sin (2\pi\phi(t)) + V_{sys} + v_{bary}(t)
\end{equation}
with
\begin{equation}\label{eq:11}
\phi(t)= \frac{t-T_{0}}{P}
\end{equation}
where $v_{bary}(t)$ is the barycentric correction, $\phi(t)$ is the planet orbital phase, $t$ is the mid observation time in BJD, $T_{0}$ is the ephemeris, and $P$ is the orbital period of the planet, $K_{p}$ is the semi-amplitude of the radial velocity of the planet, and $V_{sys}$ is the systemic velocity \footnote{The nodal precession of WASP-33b caused the orbital inclination to evolve from $\sim$ 86.61$^{\circ}$ in 2008 to 88.70$^{\circ}$ in 2014 \citep{Johnson2015ApJ...810L..23J}, and based on \citet{Smith2011MNRAS.416.2096S} analysis on its orbital eccentricity, we can safely assume a circular orbit with orbital inclination $\sim$ 90$^{\circ}$ to calculate RV$_{p}$ }. We consider the radial velocity of the planet semi-amplitude between $+$150 km s$^{-1}$ $\leq$ $K_{p}$ $\leq$ $+$310 km s$^{-1}$, and systemic velocity between $-$80 km s$^{-1}$ $\leq$ $V_{sys}$ $\leq$ $+$80 km s$^{-1}$ with 0.5 km s$^{-1}$ steps. The result is a 321 ($V_{sys}$ as column)$\times$ 320 ($K_{p}$ as row) CCF matrix. Then this matrix was divided by its the standard deviation to make $K_{p}$ -- $V_{sys}$ S/N map.

\begin{figure*}[!ht]
\centering
\gridline{\fig{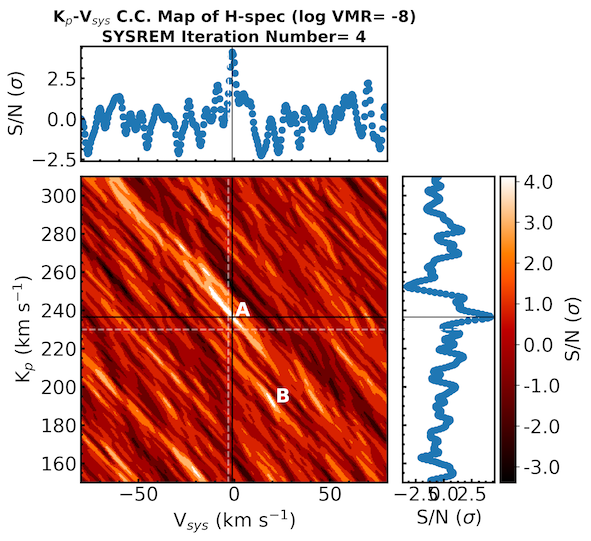}{0.48\textwidth}{(a)} 
         \fig{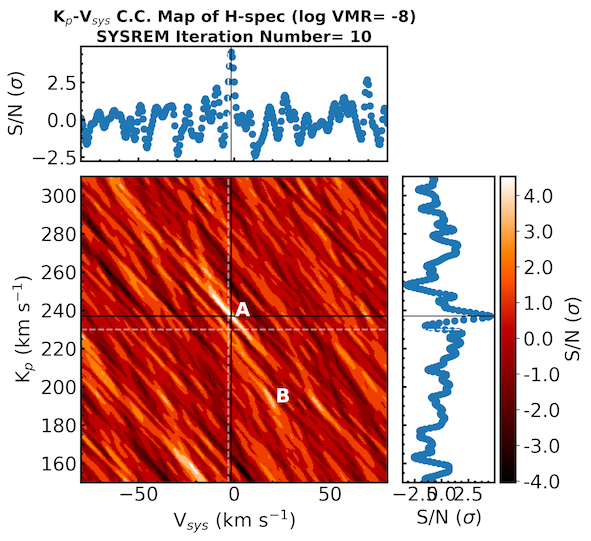}{0.48\textwidth}{(b)}}
\gridline{\fig{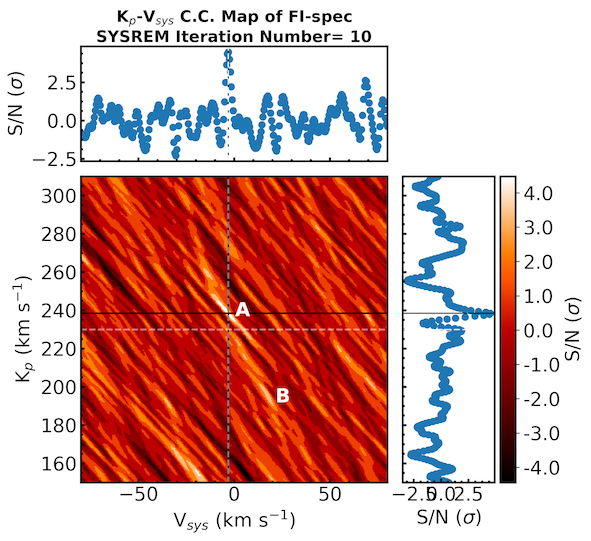}{0.48\textwidth}{(c)}
		  \fig{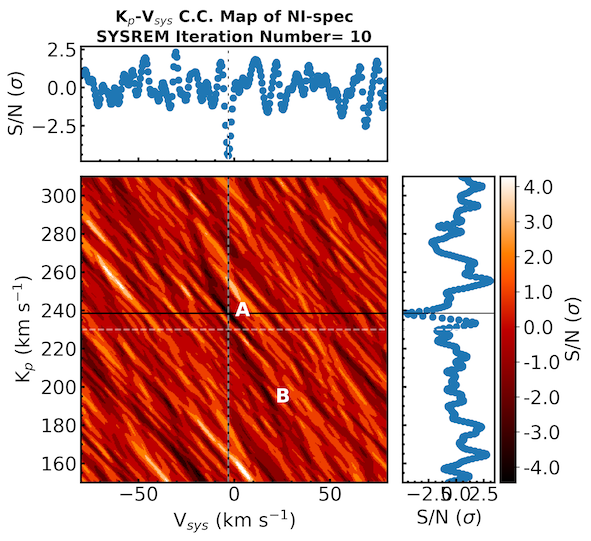}{0.48\textwidth}{(d)}}
\caption{(a) and (b) are the $K_{p}$ -- $V_{sys}$ cross-correlation map of H-spec (log VMR= -8) for SYSREM iteration number= 4 and 10 respectively. (c) and (d) are the $K_{p}$ -- $V_{sys}$ cross-correlation map of FI-spec and NI-spec for SYSREM iteration number= 10 respectively. The white dashed line is the expected Kp and Vsys from the previous studies. The black line is the maximum S/N peak in the map. The top and right panels show the 1-dimensional cross section of the CCF peak along the $V_{sys}$ and $K_{p}$ respectively. A and B are the peak A and peak B respectively. \label{fig:kpvsyssniter}}
\end{figure*}

Figure \ref{fig:kpvsyssniter} shows the $K_{p}$ -- $V_{sys}$ S/N map for $N_{\mathrm sys} = 4$ (H-spec) and 10  (H-, FI-, and NI-spec models). In all the cases except for NI-spec, we detected positive peaks with $>$ 4-$\sigma$, while the map for NI-spec exhibits a negative peak at the same place. The maps for H-spec and FI-spec exhibits the strong peak in almost all SYSREM iteration number at $K_{p}$= $+$237.0 km s$^{-1}$ and $V_{sys}$= $-$1.5 km s$^{-1}$ (henceforth peak A). Among all H-spec, the strongest signal was found in log VMR= $-$8. This peak is also the strongest one in the $K_{p}$ -- $V_{sys}$ c.c. maps of both FI-spec and NI-spec for most of the SYSREM iteration number, although for NI-spec the value is negative. The positive value of the peak A in H-spec and FI-spec, and the negative value in NI-spec suggested that non-inversion atmosphere is unlikely for WASP-33b if peak A is the real signal. The second strongest peak was detected at $K_{p}$= $+$192.0 km s$^{-1}$ and $V_{sys}$= $+$19.5 km s$^{-1}$ (henceforth peak B) in the SYSREM iteration number= 2 only. 

\begin{figure}[hbt!]
\centering
\includegraphics[width=0.45\textwidth]{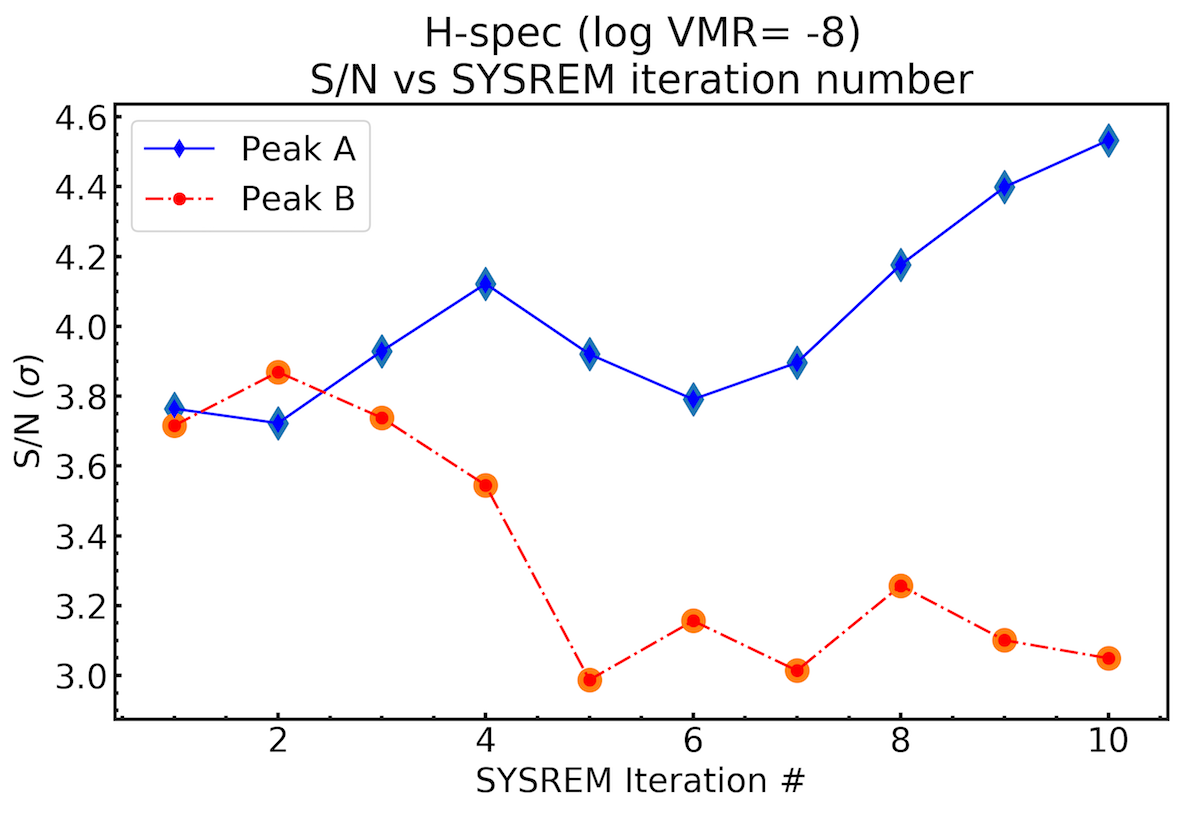}
\caption{The S/N of peak A and B along the SYSREM iteration number, $N_{\mathrm{sys}}$, for H-spec (log VMR= $-$8). The blue diamond-lined is the S/N of peak A and the red circle-dotted line is the S/N of peak B. \label{fig:sniter}}
\end{figure}

Figure \ref{fig:sniter} shows the S/N of peaks A and B as a function of the SYSREM iteration, although peak B is unlikely to be a real signal, due to its physically non-realistic $K_{p}$ and $V_{sys}$ values compared with the expected value from previous studies, it was chosen as a representative of the noise/false-positive signal. The S/N of peak B decreases as increasing of the SYSREM iteration number. The S/N of peak A increases from SYSREM iteration= 1 until SYSREM iteration= 4 then decreases until SYSREM iteration= 6 before begin to increase again until SYSREM iteration= 10. The increasing of S/N of peak A after SYSREM iteration= 6 is most likely due to the difference level of telluric and/or stellar lines removal in each order. It can be seen in the Figure \ref{fig:kpvsyssniter} a for SYSREM iteration= 4 where multiple peaks, 5 peaks, within 1$\sigma$ value from the strongest peak can be found (white color). While in the Figure \ref{fig:kpvsyssniter} b there are only 2 peaks, peak A and the other one at about $K_{p}\sim$ $+$160.0 km s$^{-1}$ and $V_{sys}\sim$ $-$20 km s$^{-1}$. From the curve of Figure \ref{fig:sniter}, it is natural to adopt $N_\mathrm{sys}=10$ as a fiducial iteration number. Because the common iteration number for all orders does not fully optimize the procedure of the systematics removal, 4.3-$\sigma$ of the significance level for H-spec (VMR= $-$8) we obtained should be regarded as a conservative estimate of detection level. 

\subsection{Order-based SYSREM Optimization \label{subsec:sysopti}}

The level of systematics (telluric lines) should be different for each order, thus to find the optimized SYSREM iteration number of each order, following steps were also performed. A scaled artificial signal was injected at the detected RV ($K_{p}$= $+$237 km s$^{-1}$ and $V_{sys}$= $-$1.5 km s$^{-1}$) in the spectra before telluric and stellar lines removal. The scaling was performed according to the equation
\begin{equation}
F_{scaled pm}(\lambda)= sc \times \frac{F_{pm}(\lambda)}{F_{star}(\lambda)} \left(\frac{R_{p}}{R_{star}}\right)^{2},
\end{equation}
where $sc$ is the scaling constant, $F_{scaled pm}$ is the scaled artificial signal, $F_{pm}(\lambda)$ is the planet model spectrum (H-spec with $\log$ VMR= $-$8) from the integration of Planck function versus monochromatic transmission along the line-of-sight (see \S \ref{sec:atmos}), $F_{star}(\lambda)$ is the black body flux of T= 7400 K representing the continuum level of the WASP-33 flux, $R_{p}$ and $R_{star}$ are the planet and star radius, respectively.

To check the dependence on the strength of the injected signal, we adopt 5 different injected signals with $sc=$ [0.2, 0.4, 0.6, 0.8, 1.0]. The injected signals are broadened by a rotation kernel, using \textit{fastRotBroad} from PyAstronomy with $v.\sin i$= 0.4 times the expected projected velocity of WASP-33b \citep[calculated by assuming a tidally locked condition with the parameter in the Table \ref{tab:table1} referring to the typical broadening width of tidally locked exoplanet as shown in][]{Kawahara2012ApJ...760L..13K}. We convolve with a box-function to take into account the change of Doppler shift during the 600 s exposures per frame. The signals were injected into the spectra, before performing telluric and/or stellar lines removal using SYSREM algorithm. The residuals for each SYSREM iteration were cross-correlated with the artificial signal itself. 

\begin{figure*}[htb!]
\centering
\includegraphics[width=0.95\textwidth]{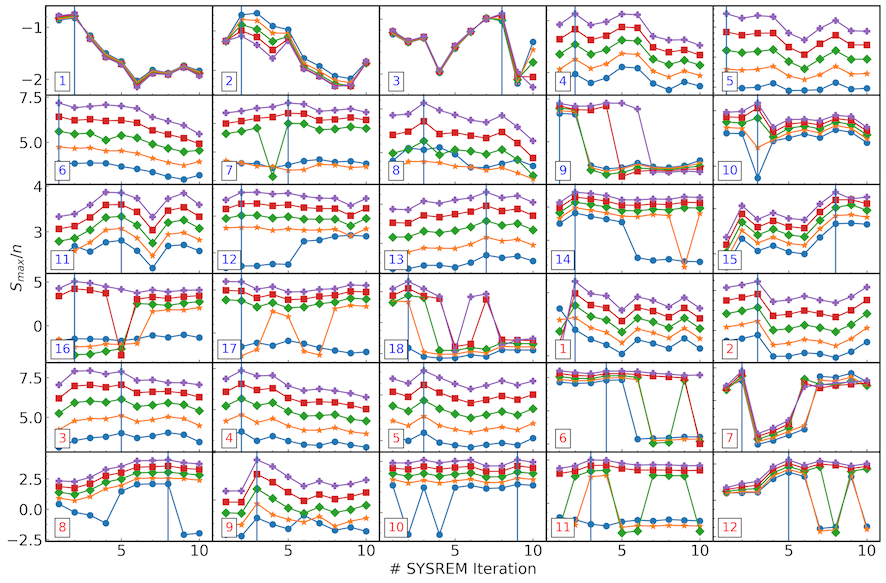}
\caption{$S_{max}/n$ of all orders for various SYSREM iteration numbers. The vertical lines mark the SYSREM iteration number that made the signal strength optimum. The blue circle, yellow asterisk, green diamond, red square, and purple cross are the $S_{max}/n$ of the recovered injected signal with $sn$= [0.2, 0.4, 0.6, 0.8, 1.0] respectively. The number in the white box at the bottom left of each panel represents the order number and the color represents the CCD (blue or red). \label{fig:sysremopt}}
\end{figure*}

The CCF of each order of each frame was aligned to the planet rest-frame according to the injected $K_{p}$ -- $V_{sys}$ (e.g. the middle panel of Figure \ref{fig:figfinalccfmap}). Then, the aligned CCFs of all frames except the frames corresponding to the secondary eclipse were summed to create the total mean CCF. To examine the exoplanet signal for various combinations of SYSREM iteration numbers, the peak value of the total mean CCF in $\pm$ 2.5 km s$^{-1}$ was taken between the expected rest-frame radial velocity of the (0 km s$^{-1}$) planet and divided by the its outside standard deviation (non-detection CCF value) to include the ``noise'' (henceforth \textit{$S_{max}/n$}). The results for five injected signals are shown in Figure \ref{fig:sysremopt}.

\begin{figure}[htb!]
\includegraphics[width=0.45\textwidth]{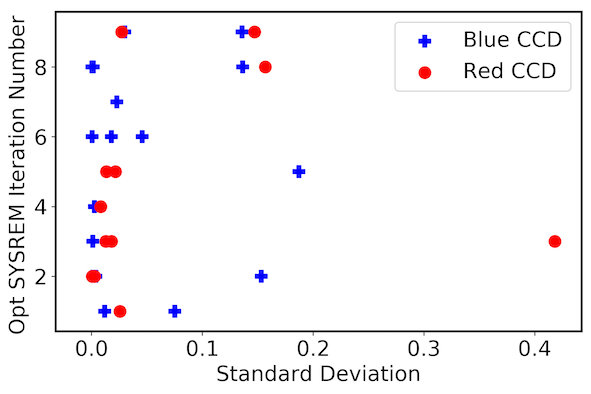}
\caption{Standard deviation of the spectrum of each order with their corresponding optimized SYSREM iteration number. The blue cross and the red circle indicate the spectrum in the blue and red CCDs, respectively. The value of the standard deviation represents the amount of telluric and/or stellar lines contamination for every order.\label{fig:figovariter}}
\end{figure}

\begin{figure}
\centering
\includegraphics[width=0.45\textwidth]{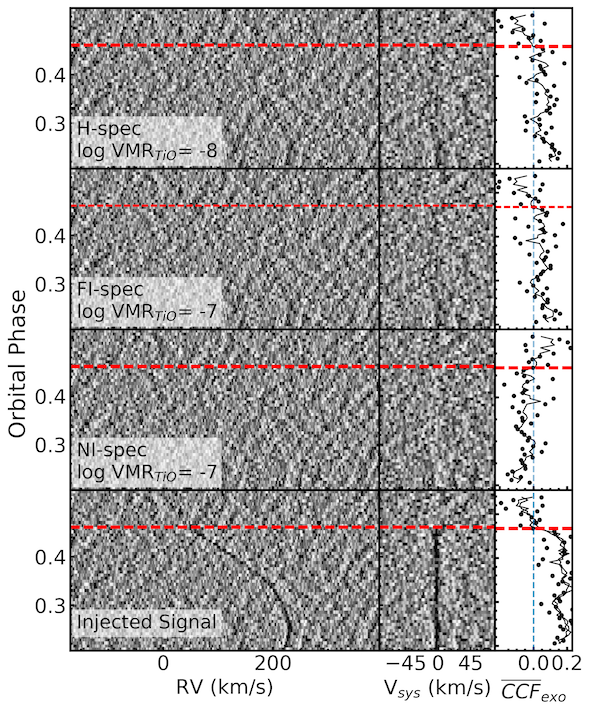}
\caption{Order-based optimized version of cross-correlation map for H-spec with log VMR= $-$8 (first row), FI-spec (second row), NI-spec (third row), and the injected signal (fourth row). The left panel shows the mean CCF map. The middle panel shows the aligned mean CCF map at $K_{p}$= $+$237.5 km s$^{-1}$ and $V_{sys}$= $-$1.5 km s$^{-1}$. The planet signal can be seen as a dark trail in the expected rest-frame of the radial velocity of the planet from the first observed phase until the appearance of the secondary eclipse phase (red dashed). The right panel shows the mean CCF for a 6 pixel column bin (CCF$_{exo}$) centered on $V_{sys}$= $-$1.5 km s$^{-1}$ along the orbital phase. The solid black line shows the smoothed CCF$_{exo}$ with a 3 frame smoothing window. The blue dashed line shows the zero value of CCF. \label{fig:figfinalccfmap}}
\end{figure}

The signal strength ($S_{max}/n$) behaves differently in every order, depending on the amount of telluric and/or stellar lines contamination and the strength of the injected signal (represented by $sc$ value). There are several jumps in $S_{max}/n$ curves, which may be caused by spurious signal, thus by following the general trend of the recovered $S_{max}/n$ curves of 5 different injected signals the optimum SYSREM iteration number we chose the optimum SYSREM iteration numbers. For blue and red CCDs the optimum SYSREM iteration numbers are [2, 2, 8, 2, 1, 1, 5, 3, 1, 3, 5, 2, 7, 2, 8, 2, 2, 2] and [2, 3, 5, 2, 3, 4, 2, 8, 3, 9, 3, 5], respectively, for each of its orders. Figure \ref{fig:figovariter} shows the level of telluric and/or stellar lines contamination of each order, which is represented by the standard deviation value of the standard reduced spectrum in that order, and its corresponding optimum SYSREM iteration number. One can see the tendency that the optimum SYSREM iteration number increases as exhibiting more telluric and/or stellar lines in the spectrum, except for one in the bottom right of the figure ( the order having many O$_{2}$ lines).

Using the optimum SYSREM iteration number, the mean CCF map was calculated for all of the spectrum model as shown in Figure \ref{fig:figfinalccfmap} (left panel). Then the CCF was aligned to the rest-frame of the planet (middle panel). The TiO signal can be seen in the left panels as a positive (dark) signal with an arc shape for H- and FI- spec and as a negative (bright) one for NI-spec. In the middle panels of the figures, the planet signal was aligned such that it can be seen as a vertical dark/bright trail at the $V_{sys}$= $-$1.5 km s$^{-1}$. The right panels show the mean CCF with the width of 6 pixels centered at $V_{sys}$= $-$1.5 km s$^{-1}$ (CCF$_{exo}$). For FI- and H-models, the CCF curve exhibits a positive offset during the visible phase while there is no offset when the planet is behind the star. This feature supports the atmospheric origin of the CCF signal. The exposure time for a single frame is 600 s, which corresponds to $\sim$ 7 km s$^{-1}$ at the near-eclipse phase. This long exposure time should have smeared out the day-side spectrum by the change in the radial component of the orbital velocity of the planet, especially in the phases near the secondary eclipse. This smearing effect may account for the fact that the signal is only visible at the phase $\lesssim$ 0.35, while it should be stronger at the phase $\gtrsim$ 0.35 because larger part of the dayside of the WASP-33b is visible.  

\begin{figure}[t]
\centering
\includegraphics[width=0.45\textwidth]{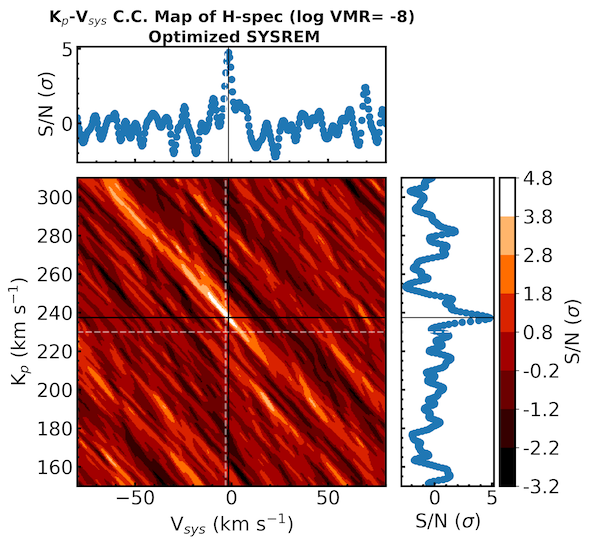}
\caption{Figure $K_{p}$ -- $V_{sys}$ S/N map of H-spec with maximum peak at $K_{p}$= $+$237.5 $^{+13.0}_{-5.0}$ km s$^{-1}$ and $V_{sys}$= $-$1.5 $^{+4.0} _{-10.5}$ km s$^{-1}$, which gives 4.8-$\sigma$ detection. The white dashed line is the expected Kp and Vsys from the previous studies. The top and right panels show the 1-dimensional cross section of the CCF peak along the $V_{sys}$ and $K_{p}$ respectively. The black dash lines show the most significant signal, the white dash lines show the expected $K_{p}$ and $V_{sys}$, and the color bar grid interval is 1-$\sigma$, the white area also represents the 1-$\sigma$ error of the detected signal.\label{fig:finopthay8}}
\end{figure}

Figure \ref{fig:finopthay8} shows a $K_{p}$ -- $V_{sys}$ S/N map for H-spec with $\log$ VMR$_{TiO}$= $-$8. The noise level in the $K_{p}$ -- $V_{sys}$ S/N map after optimization is significantly suppressed. The strongest peak was found with 4.8-$\sigma$ detection significance at $K_{p}$= $+$237.5 $^{+13.0}_{-5.0}$ km s$^{-1}$ and $V_{sys}$= $-$1.5 $^{+4.0} _{-10.5}$ km s$^{-1}$ within an elliptic region of 1 sigma. The latter is consistent with the estimates from the stellar RV \citep[\S \ref{subsection:wasp33rv} and ][]{Collier2010MNRAS.407..507C}. Because we already have the strong constraint on $V_{sys} = -3.02 \pm 0.42$ km s$^{-1}$ from the stellar spectrum (\S \ref{subsection:wasp33rv}), we may assume this value as the prior of $V_{sys}$ on the $K_{p}$ -- $V_{sys}$ S/N map. Then we obtain a stronger constraint, $K_{p} = 239.0^{+2.0}_{-1.0}$ km s$^{-1}$. This is the first time that the orbital velocity of WASP-33b was dynamically measured. The orbital velocity of a planet depends on the orbital period and the mass of the host star. Since the measurement of orbital period is very precise for a transiting exoplanet, our observation provides the first model-independent measurement of the mass of the host star. Using the period of WASP-33b from the previous study (see Table \ref{tab:table1}) the measured mass of the host star is $M_{\star}= 1.73 ^{+0.04}_{-0.02} $ $M_{\bigodot}$, which is larger than values previously reported.

\subsection{Statistical Tests} \label{subsec:statanal}
\begin{figure}[t!]
\centering
\includegraphics[width=0.45\textwidth]{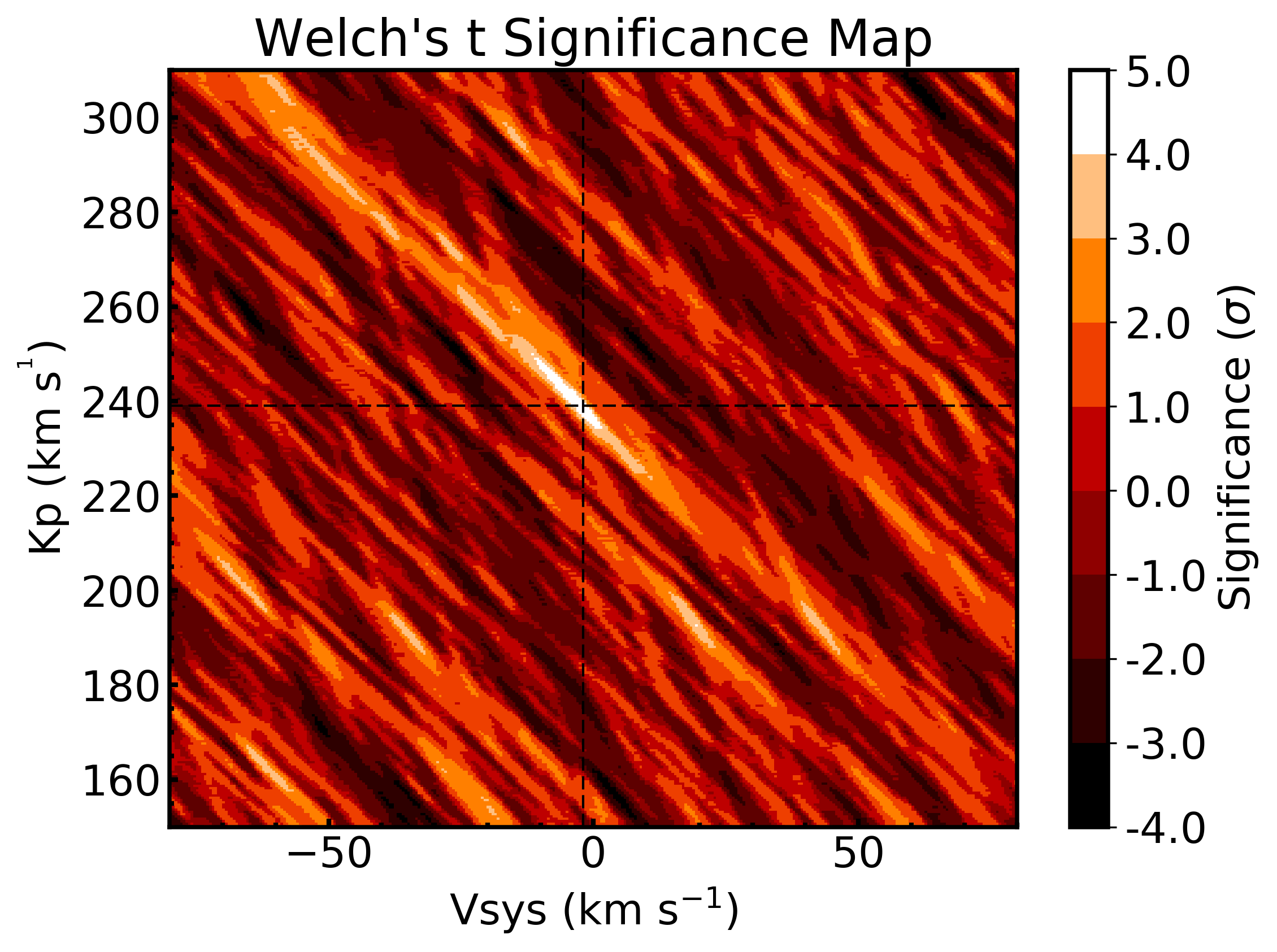}
\caption{Significance map after converting from the $p$-value from Welch's \textit{t}-test using $erf$ function for the most significant detected signal with the H-spec model spectrum of $\log$ VMR$_{TiO}$= $-$8. The black dashes show the most significant signal, the white dashes show the expected $K_{p}$ and $V_{sys}$, and the color bar grid interval is 1-$\sigma$, the white area also represents the 1-$\sigma$ error of the detected signal. \label{fig:welch}}
\end{figure}

\begin{figure}[]
\centering
\includegraphics[width=0.45\textwidth]{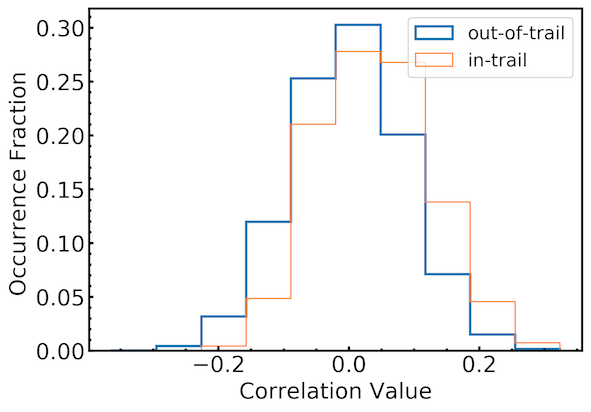}
\caption{Histogram of the in-trail and out-trail mean CCF distribution, the in-trail distribution is slightly shifted from the out-of-trail distribution.\label{fig:fighistmax}}
\end{figure}

\begin{figure}[]
\centering
\includegraphics[width=0.4\textwidth]{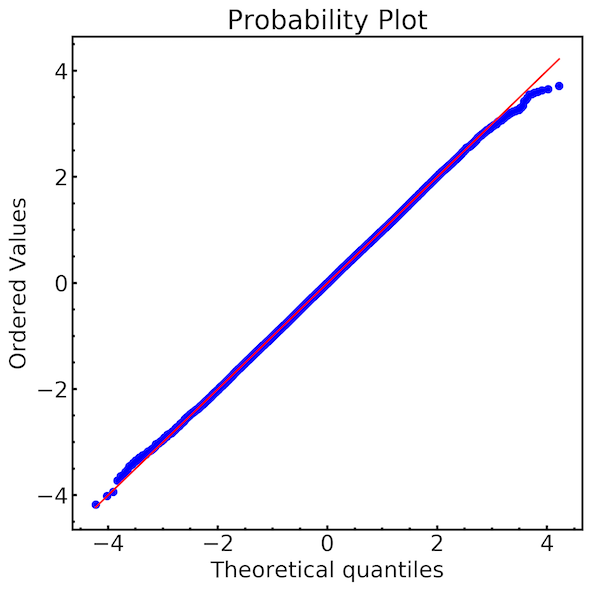}
\caption{Q-Q plot of the out-of-trail distribution which shows that the distribution is Gaussian until 4-$\sigma$. \label{fig:figprobmax}}
\end{figure}

The in-trail CCF (henceforth in-trail signal) was compared with the out-of-trail CCF (henceforth out-of-trail signal) by performing Welch's \textit{t}-test to check if the mean is same assuming that two distributions were drawn from the same parent distribution, using the \textit{SciPy} module in Python 2.7. We used equation (\ref{eq:10}) to calculate the expected RV$_{p}$ for the same range to ones for the $K_{p}$ -- $V_{sys}$ CCF maps, and took the 1 pixel mean CCF value at the closest RV to RV$_{p}$. The out-of-trail signal contains all mean CCF values except the in-trail signal. 

The out-of-trail and in-trail signal histograms (width of the in-trail signal= $\pm$ 3 pixels) were plotted. The in-trail signal distribution shifted further from the out-of-trail signal distribution. The distribution of the in-trail signal is clearly shifted from the zero values. From the Q-Q plot (see Figure \ref{fig:figprobmax}) it can be seen that the out-of-trail signal distribution is a Gaussian distribution until 4-$\sigma$, therefore we can safely convert the half $p$-value to $\sigma$ value of the detection significance using an error function\footnote{The output from the \textit{scipy} module is the two-tailed $p$-value.}. The Welch's \textit{t} test shows that the in-trail distribution is deviated from the out-of-trail signal distribution by 5.0-$\sigma$ (see Figure \ref{fig:welch}) in line with the S/N of this peak detection.

\section{Discussion and Conclusions}\label{sec:discussandcon}

We confirmed the previously claimed of the inaccuracy of the TiO line list for wavelengths shorter than 6300 $\AA$ \citep[see][] {Hoej2015A&A...575A..20H}. We also showed that the line list accuracy is enough high for longer wavelengths, which was considered in processing order-based optimization, therefore our analysis no longer suffers from the inaccuracy issue. By measuring the radial velocity of the host star we also confirmed the measurement by \citet{Collier2010MNRAS.407..507C}, which differs by $\sim$ 4 km s$^{-1}$ from the SIMBAD database. This confirmation gives us a narrow $V_{sys}$ search space to find the possible exoplanet signal.

We reported a TiO molecule signature detection in the day-side spectra of WASP-33b by 4.8-$\sigma$ confidence level, which provided direct evidence of the existence of TiO in the atmosphere of the hot Jupiter. The detection levels for VMR= -8, -9, and -10 (H-spec) lie within 1-$\sigma$ from the highest one (VMR= -8); moreover, in our analysis we did not use a self-consistent atmospheric model, thus the constraint on the VMR cannot be obtained directly from our result only. Our TiO molecular detection for H-spec and FI-spec confirmed the existence of stratosphere (thermal inversion layer) in the day-side of WASP-33 b, as previously has been claimed by several studies \citep[e.g.][]{2015A&A...584A..75V, Haynes2015ApJ...806..146H}. The full inversion T/P profile has also been reported for another super hot Jupiter, WASP-121b by \cite{Evans2017Natur.548...58E}, which resolved the emission spectral feature of H$_{2}$O at near-infrared wavelength, using HST. Our result is complementary to the TiO detection using low-dispersion spectroscopy in WASP-19b by \citet{2017Natur.549..238S}, who were able to constrain the relative abundance of TiO (0.12 p.p.b) but could not provide information about T/P profile as they were only able to measure the transmission level of the molecules in the atmosphere. The constraint on the relative abundance level of TiO in WASP-33b can be obtained by analyzing it using a self-consistent atmospheric model and/or by combining low- and high-dispersion spectroscopy in order to introduce a more precise constraint of the relative abundance of each detected molecules and the T/P profile of the exoplanet atmosphere \citep{2017ApJ...839L...2B}.

By observing for about nine hours only using the 8.2 m Subaru telescope, we are able to detect significant signature of TiO in the atmosphere of WASP-33b. Our results demonstrate that high-dispersion spectroscopy is a powerful tool to characterize the atmosphere of an exoplanet, and show a promising potential of developing similar/more advanced techniques for the Infra Red Doppler instrument \citep{2014SPIE.9147E..14K} in the Subaru telescope, and for extra large telescopes facilities in the future, as suggested by several authors \citep[e.g.][]{2014arXiv1409.5740K,2015A&A...576A..59S}.

\section{Acknowledgement} \label{sec:acno}
This work was based in part on data collected on HDS at the Subaru Telescope, which is operated by the National Astronomical Observatory of Japan. We thank our anonymous referee, whose insightful comments improved the manuscript. S.K.N. acknowledges support from Indonesia Endowment Fund for Education Scholarship. S.K.N also acknowledges Toru Yamada for fruitful discussions during the analysis of the data. H.K. is supported by Grant-in-Aid for Young Scientist (B) from Japan Society for Promotion of Science (JSPS), No. 17K14246 and the Astrobiology Center from NINS, No. AB291003. T.H. is supported by Grant-in-Aid for Young Scientist (B) from Japan Society for Promotion of Science (JSPS), No. 16K17660.
\linebreak

\appendix

\section{Correction for Blaze Function Variation and Normalization of Spectra \label{Appendix_A}}
\begin{figure*}[h]
\includegraphics[width=1\textwidth]{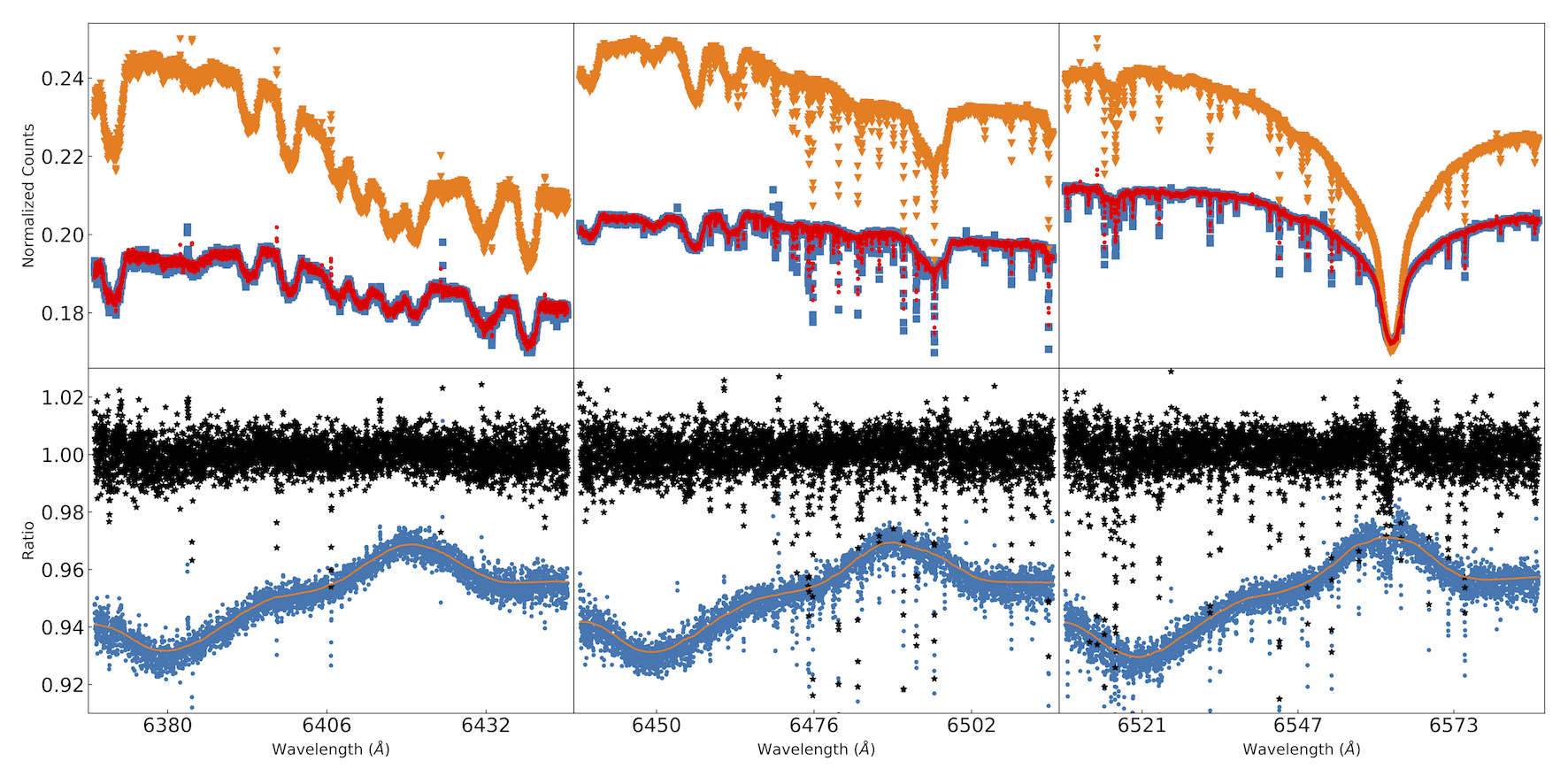}
\caption{Top panel: Spectra from two different epoch for order 93 (left panel), 92 (middle panel), 91 (right panel), which were taken at the beginning (blue box) and the end (orange triangle) of the observation. The continuum profile of the second spectrum after correction (red circle) matched with the first spectrum. Bottom panel: ratio of the spectra before (blue circle) and after (black star) correction. The orange line is the smoothed ratio profile used to correct the variation. \label{fig:figblaze}}
\end{figure*}

As \citet{Winn2004PASJ...56..655W} showed, observation using HDS suffered from variation of the blaze function during the observation. The continuum profile is not important in our analysis, but it is useful to correct this variation as a method of subtracting the continuum of each spectrum. We used the first frame as a reference spectrum and compared the rest of spectra by calculating the ratio of both spectra. By looking at the ratio of all spectra, we found wavelength variations with similar patterns for all orders, but these vary along the time of the observation (see Figure \ref{fig:figblaze}). The ratio spectra were then clipped to remove any outlier. In order to avoid any residual broad stellar absorption lines mismatch in the ratio spectra, we selected a 301-pixels smoothing window and applied smoothing, using \textit{PyAstronomy.pyasl.smooth}\footnote{\url{http://www.hs.uni-hamburg.de/DE/Ins/Per/Czesla/PyA/PyA/index.html}} with a flat window function. Each of the compared spectra was then divided by its smoothed ratio spectra, resulting in a spectra with shared blaze function with the reference spectrum, while conserving stellar lines and telluric line strength variations. A spline function was fitted to the reference spectrum using $\textit{continuum}$ manually and divided all spectra by it to get the normalized spectra.

\section{SYSREM algorithm \label{Appendix_B}}
As in \citet{Tamuz2005MNRAS.356.1466T}, the aim is to find the two sets of effective extinction coefficients ($c_{i}$) and air mass ($a_{j}$) that optimally describe the atmospheric absorption in each wavelength bin ($i$) of each frame ($j$). By taking the observed air mass as the first input of $a$, we search for the optimal $c_{i}$ that minimizes 
\begin{equation}
R_{i}^{2}= \sum_{j} \frac{\left(r_{ij}-c_{i}a_{j}\right)^{2}}{\sigma^{2}_{ij}}
\end{equation}
where $\sigma$ is the uncertainty of pixels $ij$ calculated by taking the root sum square of the standard deviations of its frame, and the wavelength bin. Then, $c_{i}$ can be estimated by
\begin{equation}
c_{i}= \frac{\sum_{j}\left(r_{ij}a_{j}/\sigma^{2}_{ij}\right)}{\sum_{j}\left(a_{j}^{2}/\sigma^{2}_{ij}\right)}
\end{equation}
Then, by using the estimated $c_{i}$, the ``optimized air mass'' ($a_{j}^{(1)}$) can also be found that minimizes
\begin{equation}
R_{j}^{2}= \sum_{i} \frac{\left(r_{ij}-c_{i}a_{j}\right)^{2}}{\sigma^{2}_{ij}}
\end{equation}
by calculating
\begin{equation}
a_{j}^{(1)}= \frac{\sum_{i}\left(r_{ij}c_{i}/\sigma^{2}_{ij}\right)}{\sum_{i}\left(c_{i}^{2}/\sigma^{2}_{ij}\right)}
\end{equation}
By using the ``optimized air mass'', the ``optimized coefficient'' $c_{i}^{(1)}$ can be calculated and by performing this ``criss-cross'' iteration the stable value of $\hat{c}_{i}^{(1)}$, $\hat{a}_{j}^{(1)}$ can be found. The residual can be calculated as
\begin{equation}
r_{ij}^{(1)}= r_{ij}-\hat{c}_{i}^{(1)}\hat{a}_{j}^{(1)}
\end{equation}
At this point, the first systematic effect has been removed; then, by performing a similar calculation to find $c_{i}^{(2)}$, $a_{j}^{(2)}$ that minimize 
\begin{equation}
R_{i}^{2}= \sum_{j} \frac{\left(r_{ij}^{(1)}-c_{i}^{(2)}a_{j}^{(2)}\right)^{2}}{\sigma^{2}_{ij}}
\end{equation}
the second and subsequent systematics can be calculated and removed. Note that $\hat{c}_{i}\hat{a}_{j}$ does not actually represent the real extinction coefficient and air mass even for the first iteration, but a linear systematic effect that varies as a function of wavelength and time (or frame number).

\section{Cross section of molecules\label{Appendix_C}}

The TiO line list from \citet{1998A&A...337..495P} was used, which include five different isotopes with nine electronics systems, and the partition function that was published by \cite{2016A&A...588A..96B}. As Py4CATS only supports HITRAN- and GEISA-like databases, instead of using its \textit{extract} module, we used our custom build Python script to extract the TiO line list and wrote them in HITRAN-like format, which then were used to calculate line-by-line cross-sections. The cross-section ($k$) of each line is a product of the line strength ($S$) and a normalized line profile function ($g$)
\begin{equation}\label{eq:1}
k(\nu;\hat{\nu}, S, \gamma)= S(T) \ \cdot \ g(\nu;\hat{\nu}, \gamma) \ \text{with} \ \int_{-\infty}^{\infty}g\ \text{d}\nu=1
\end{equation}
where $\gamma$ is the line broadening half width at half maximum (HFWHM), $\nu$ is the frequency, and $\hat{\nu}$ is the line centroid position. We modified \textit{lbl2xs.py} at the adjusting line parameter of the $p$ and $T$ section to enable it to calculate line strength using other partition function databases, and to include thermal (Doppler), natural, and van der Waals broadening for TiO in the line profile calculations. The line strength at temperature T ($S(T)$) is calculated by using Equation (1) in \citet{Sharp2007ApJS..168..140S}. A Voigt function was used for the line profile, which is defined as
\begin{equation}\label{eq:2}
K(x,y)= \frac{y}{\pi} \int_{-\infty}^{\infty}\frac{e^{-t^{2}}}{(x-t)^{2}+y^{2}}dt
\end{equation}
where $x,y$ are defined as dimensionless variables in terms of distance from the line centroid position, $\nu-\hat{\nu}$, and the ratio of Lorentz and Gaussian  HWHM $\gamma_{L}$, $\gamma_{D}$:
\begin{equation}\label{eq:3}
x=\sqrt{\ln\ 2}\ \frac{\nu-\hat{\nu}}{\gamma_{D}} \quad \text{and} \quad y=\sqrt{\ln\ 2}\ \frac{\gamma_{L}}{\gamma_{D}}
\end{equation}
The thermal broadening is expressed with a Gaussian profile ($g_{D}$):
\begin{eqnarray}\label{eq:4}
g_{D}(\nu)= \frac{1}{\gamma_{D}}\ \left(\frac{\text{ln}\ 2}{\pi}\right)^{1/2}\text{exp} \left[ - \text{ln}\ 2 \left( \frac{\nu-\hat{\nu}}{\gamma_{D}}\right)^{2}\right] \\
\text{with} \ \ \gamma_{D} = \hat{\nu} \sqrt{\frac{2\ \text{ln}\ 2 kT}{mc^{2}}} \nonumber
\end{eqnarray}
while natural and van der Waals broadening is expressed with a Lorentz profile ($g_{L}$)
\begin{equation}\label{eq:5}
g_{L}(\nu)= \frac{\gamma_{L}/\pi}{(\nu-\hat{\nu})^{2}+\gamma_{L}^{2}}
\end{equation}
where $k$ is Boltzmann's constant, $T$ is temperature, $m$ is the mass of the molecular absorber, and $c$ is the speed of light. $\gamma_{L}$ is a sum of van der Waals ($\gamma_{LW}$) and natural ($\gamma_{LN}$) line broadening HWHM,
\begin{equation}\label{eq:6}
\gamma_{L} = \gamma_{LW} + \gamma_{LN}
\end{equation}

As there was no information available for the van der Waals line broadening width of both molecules, we calculated it by following \citet{Sharp2007ApJS..168..140S} as
\begin{equation}\label{eq:7}
\gamma_{LW} = \gamma_{0} \frac{p}{p_{0}} \times \left( \frac{T_{0}}{T} \right)^{n}
\end{equation}
with 
\begin{equation}\label{eq:8}
\gamma_{0} = \frac{1}{2}     [w_{0}-min(J\arcsec,30)w_{1}]
\end{equation}
where $J\arcsec$ is the lower rotational quantum number of each energy transition, $w_{0}$ is the FWHM of a transition at 1 atm ($p_{0}$) when $J\arcsec = 0$, $w_{1}$ is a scale factor of the dependency of the broadening on $J\arcsec$, and as suggested in the paper we used $w_{0}= 0.1$ cm$^{-1}$, $w_{1}= 0.002$ cm$^{-1}$, and $n=$ 0. The minimum criterion in Equation $\ref{eq:8}$ means that the line broadening at $J\arcsec = 30$ is used for larger $J\arcsec$ values. We used the formalism by Gray (1976) for the natural broadening width:
\begin{equation}\label{eq:9}
 \gamma_{LN} = \frac{0.222\ \hat{\nu}^{2}}{4\pi c}
\end{equation}

\end{document}